\documentclass[showpacs,twocolumn,prl,aps]{revtex4}
\usepackage{amsmath}
\usepackage{graphicx}
\usepackage{bm}

\begin{document}

\title[Short title for running header]{Competing emergent Potts orders and possible nematic spin liquids in the Kagome $J_{1}-J_{3}$ Heisenberg model}
\author{Hong Li and Tao Li}
\affiliation{Department of Physics, Renmin University of China, Beijing 100872, P.R.China}
\date{\today}

\begin{abstract}
The extreme frustration in the Kagome antiferromagnet makes it possible to realize a large number of closely competing magnetic phases either at the classical or the quantum level.  Motivated by recent neutron scattering study on the Kagome antiferromagnet vesignieite BaCu$_{3}$V$_{2}$O$_{8}$(OH)$_{2}$, we have made systematic investigation of the phase diagram of the classical and quantum $J_{1}-J_{3}$ antiferromagnetic Heisenberg model defined on the Kagome lattice(KAFHM). While it is shown previously that a large antiferromagnetic exchange between the third-neighboring spins can drive an emergent  $q=4$ Potts order through the order-by-disorder mechanism, it is elusive what will happen in the so called "grey region" where the Luttinger-Tisza criteria fails to predict the classical ordering pattern. Through extensive Monte Carlo simulation, we find that the "grey region" is characterized by competing emergent $q=3$ and $q=4$ Potts order, whose emergence are beyond the description of the conventional order-by-disorder mechanism. Our Schwinger Boson mean field calculation in the "grey region" confirms the existence of the Potts-3 order in the ground state of quantum $J_{1}-J_{3}$ KAFHM. The predicted Potts-3 state is found to take two different forms distinguished by their projective symmetry group character(PSG or quantum order), although they have exactly the same symmetry. The transition between these two quantum orders is found to occur at a $J_{3}$ value intriguingly close to the trictritical point in the Potts-3 phase of the classical phase diagram of the model. A comparison with the exact diagonalization result on a 36-site cluster shows that the "grey region" may indeed host a nematic spin liquid ground state featuring an anisotropic ring structure around the $\mathbf{q}=0$ point in its spinon dispersion relation. We find that such "grey region" of extremely frustrated magnets should better be taken as the playground to study the rich competition between exotic emergent phases, rather than a burden on their theoretical analysis.      
\end{abstract}

\pacs{}

\maketitle

\section{I. Introduction}
The Kagome antiferromagnetic Heisenberg model(KAFHM) exhibits a very rich competition of magnetic phases both at the classical and the quantum level as a result of extreme geometric frustration on the Kagome lattice. At the classical level, the KAFHM with only first-neighboring exchange possesses an extensive ground state degeneracy, setting the stage for the competition of a large number of symmetry breaking phases when one turn on small additional exchange couplings. Such a singular situation is not fully resolved even if we include quantum fluctuation correction. For example, while the ground state of the spin-$\frac{1}{2}$ KAFHM with first-neighboring exchange is widely believed to be a quantum spin liquid state\cite{ED1,ED2,ED3,HTSE1,ED4,ED5}, its exact nature is still under debate\cite{HTSE2,TRG1,DMRG1,DMRG2,DMRG3,DMRG4,VMC1,VMC2,VMC3,VMC4,VMC5,VMC6,TRG2}. Accompanying this debate is the mystery on the origin of the  massive number of spin singlet excitation below the tiny spin triplet gap as found in exact diagonalization studies\cite{Singlet1,Singlet2,Singlet3,Singlet4}. Such exotic behaviors have motivated several theoretical suggestions that the spin-$\frac{1}{2}$ KAFHM with first-neighboring exchange may sit at or be very close to a quantum critical point\cite{QCP1,QCP2,QCP3}, where two or even a massive number of phases meet. The massive number of spin singlet excitation below the tiny spin triplet gap may just serve as the bridge interpolating between the closely competing phases. Thus, the study of perturbation away from the spin-$\frac{1}{2}$ KAFHM with first-neighboring exchange can provide an illuminating way to elucidate the exotic physics of this particular model\cite{DMRG5,DMRG6,DMRG7,DMRG8,Changlani1,Changlani2}. Such perturbation also appears naturally in various material realization of the KAFHM such as kapellasite\cite{Kap}, volborthite\cite{Vol}, haydeite\cite{Hay1,Hay2}, Ba-vesignieite\cite{Bav1,Bav2,Bav3,Bav4,Bav5}, and Sr-vesignieite\cite{Srv} and is thus of great experimental importance.

The consequence of the geometric frustration in a quantum magnet can often be learned from an analysis of the corresponding classical model. For a general Heisenberg model of the form $H=\sum_{i,j}J_{i,j}\mathbf{S}_{i}\cdot\mathbf{S}_{j}$ defined on a simple Bravais lattice, the semiclassical ordering pattern of the system can be determined by the so called Luttinger-Tisza criteria\cite{Tisza}. More specifically, the ordering wave vector is given by the momentum at which the Fourier transform of the exchange couplings $J(\mathbf{q})$ reaches its minimum. For system with multiple minimum in $J(\mathbf{q})$,  the remaining degeneracy in the semiclassical ground state can usually be lifted by the conventional order-by-disorder mechanism from either thermal or quantum fluctuations. Such fluctuation correction usually favors collinear magnetic ordering pattern. The most well known example of such an order-by-disorder mechanism being the emergence of stripy magnetic order in the $J_{1}-J_{2}$ antiferromagnetic Heisenberg model defined on the square lattice(SAFHM). 

For an non-Bravais lattice, namely a lattice with more than one site per unit cell, $J(\mathbf{q})$ becomes a matrix. We can still use the eigenvalue of  $J(\mathbf{q})$ as an indicator of the semiclassical ordering pattern\cite{Tao1}. More specifically, the minimum with respect to $\mathbf{q}$ of the lowest eigenvalue of $J(\mathbf{q})$ provides a lower bound on the semiclassical ground state energy. When a trial spin configuration reaches this lower bound it is then within the degenerate ground state manifold. However, the Luttinger-Tisza criteria is in general invalid as the eigenvectors of $J(\mathbf{q})$ usually violates the constraint of a uniform length for the local moment. When the lower bound provided by the Luttinger-Tisza analysis can not be reached by any trial spin configuration, there will be no general way to determine the classical ground state of the system. We will call such parameter region as "grey region" below. Obviously, the "grey region" is the most susceptible to the effect of quantum fluctuation and thus holds the best opportunity to realize the quantum spin liquid ground state, while in many previous studies such strongly frustrated region is left untouched for technical reason. 

According to the Mermin-Wagner theorem, spontaneous breaking of the continuous spin rotational symmetry is prohibited at any finite temperature in two space dimension. Thus a finite temperature transition can only occur through the breaking of a discrete symmetry for a general two dimensional frustrated magnet in a disordered spin background. The discrete symmetry that can be spontaneously broken includes the lattice translation symmetry, the lattice rotation or inversion symmetry, the time reversal symmetry, and different combinations of them. These discrete symmetry breaking orders can be driven by the conventional order-by-disorder mechanism when the Luttinger-Tisza criteria applies. As we will show in this work, such discrete symmetry breaking order can also emerge in the mysterious "grey region" in which the Luttinger-Tisza criteria does not apply. The emergence of the discrete symmetry breaking order in such "grey region" is obviously beyond the description of the conventional order-by-disorder mechanism and thus contains much richer possibilities.

The discrete symmetry breaking of the classical model at finite temperature may be inherited by the quantum model at zero temperature with the spin of the system remaining disordered, provided that the frustration is sufficiently strong, for example, when the system is located in the "grey region". This will result in the valence bond crystal phase, the nematic spin liquid phase, and the chiral spin liquid phase, depending on which specific discrete symmetry(or combination there of) is spontaneously broken. However, different from its classical counterpart, the quantum liquid phase can have different quantum orders even if they exhibit exactly the same symmetry breaking pattern. Such distinction in the quantum order can be classified in the projective symmetry group(PSG) scheme\cite{psg,psgb,psgc}, which correspond to the difference in the symmetry fractionalization pattern of the spinon excitation is the resultant spin liquid phases. This adds an additional layer of richness in the emergence phenomena in such strongly frustrated models.

In this work, we show that all these general considerations are elegantly illustrated in the $J_{1}-J_{3}$ KAFHM with antiferromagnetic third-neighboring exchange $J_{3}$ and ferromagnetic first-neighboring exchange $J_{1}$. Such a model is believed to be of direct relevance for the description of the Kagome antiferromagnet Ba-vestigenate\cite{Bav5}. When $J_{3}$ dominates, the model is essentially composed of three decoupled sublattices of antiferromagnetic correlated spins(3sub-AF). The effect of the first-neighboring exchange is exactly cancelled at the classical level in such a 3sub-AF background. This is just alike the situation in the $J_{1}-J_{2}$ SAFHM with $J_{2}>J_{1}/2$. A collinear spin ordering pattern will be favored through the order-by-disorder mechanism and as a result the lattice symmetry will be spontaneously broken. More specifically, for $J_{3}/|J_{1}|>\frac{1+\sqrt{5}}{4}\approx 0.809$, when the Luttinger-Tisza criteria predicts the 3sub-AF ordering pattern in the classical ground state, a finite temperature transition toward a phase with an emergent  $q=4$ Potts order is found in recent Monte Carlo simulation of the $J_{1}-J_{3}$ KAFHM\cite{Grison}.  However, for smaller value of $J_{3}$, there appears a "grey region" ($J_{3}/|J_{1}|\in (\frac{1}{4},\frac{1+\sqrt{5}}{4})$) in the phase diagram where the Luttinger-Tisza criteria fails to predict the classical ordering pattern. Monte Carlo simulation shows that there is also a finite temperature transition in this "grey region", toward a phase with an unidentified nature for $J_{3}/|J_{1}|<0.69$ and the  $q=4$ Potts phase for $J_{3}/|J_{1}|>0.69$. It seems that the $q=4$ Potts order can penetrate smoothly into the "grey region". However, the exact nature of the phase boundary between the $q=4$ Potts phase and the phase with the unidentified order for $J_{3}/|J_{1}|<0.69$ remains elusive. In particular, it is not clear what is the role of the upper boundary of the "grey region" in the classical phase diagram.
 
With extensive Monte Carlo simulation adopting the heat bath algorithm and the over relaxation technique\cite{Young,Okubo}, we have mapped out the full classical phase digram of the $J_{1}-J_{3}$ KAFHM. We find that the $q=4$ Potts phase is separated from an emergent $q=3$ Potts phase in the "grey region" by an almost vertical first order line at $J_{3}/|J_{1}|\approx0.69$. The upper boundary of the "grey region" predicted by the Luttinger-Tisza criteria thus plays no role in determining the classical phase diagram of the system. We then use the Schwinger Boson mean field theory(SBMFT) to work out the ground state phase diagram of the quantum $J_{1}-J_{3}$ KAFHM in the "grey region". In accord with the classical phase diagram, we find the "grey region" is split into the Potts-3 part and the Potts-4 part in the ground state phase diagram predicted by the SBMFT. However, different from its classical counterpart, we find that there are two nematic spin states in the Potts-3 region with exactly the same symmetry but different PSG characters(or different quantum orders). Intriguingly, we find that the transition point between these two nematic spin states is very close to the tricritical point along the phase boundary of the Potts-3 phase in the classical phase diagram($J_{3}/|J_{1}|\approx0.4$). Our exact diagonalization calculation on a 36-site cluster shows that only the nematic spin state below $J_{3}/|J_{1}|\approx0.4$ may be stabilized, while the nematic spin state above it may be swallowed by the 3sub-AF phase as a result of the additional quantum fluctuation correction not taken into account in the semiclassical treatment. We propose the nematic spin state realized in the "grey region" to be a possible nematic spin liquid state featuring a spinon dispersion with an anisotropic ring around the $\mathbf{q}=0$ point, namely, an anisotropic analog of the double spiral phase proposed for the doped t-J model three decades ago\cite{Kane}. 

The paper is organized as follows. In Sec.II, we will introduce the $J_{1}-J_{3}$ KAFHM and sketch its classical phase diagram. In Sec.III, we will introduce the order parameters for the emergent $q=3$ and $q=4$ Potts order in the "grey region". In Sec.IV, we will introduce the heat bath algorithm and the over relaxation technique adopted in our Monte Carlo simulation. We will then present the full classical phase digram of the $J_{1}-J_{3}$ KAFHM, followed by a detailed analysis of each phase in the phase diagram. In Sec.V, we will introduce the Schwinger Boson mean field theory for the ground state of the quantum $J_{1}-J_{3}$ KAFHM and the unrestricted mean field search technique used to find the ground state phase digram of the system. We then present the SBMFT phase diagram of the model in the "grey region". This is followed by a detailed analysis of each quantum phase in the SBMFT phase diagram, especially their PSG character and spinon dispersion. Finally, we compare both the classical phase diagram and the SBMFT phase digram to the result of exact diagonalization calculation on a 36-site cluster, base on which we argue the existence of a nematic spin liquid state in the "grey region" of the quantum $J_{1}-J_{3}$ KAFHM. In Sec.VI, we will summarize our study on the $J_{1}-J_{3}$ KAFHM and draw some general conclusions on the study of frustrated magnet possessing a "grey region".

\section{II. The $J_{1}-J_{3}$ KAFHM and a sketch of its classical phase diagram}

The model we focus on in this study is the $J_{1}-J_{3}$ KAFHM defined as
\begin{equation}
H=J_{1}\sum_{\langle i,j \rangle}\mathbf{S}_{i}\cdot\mathbf{S}_{j}+J_{3}\sum_{\langle\langle\langle i,j \rangle\rangle\rangle}\mathbf{S}_{i}\cdot\mathbf{S}_{j}
\end{equation}
Here $\langle i,j \rangle$ and $\langle\langle\langle i,j \rangle\rangle\rangle$ denote the first-neighboring and the third-neighboring pairs of sites on the Kagome lattice, as is illustrated in Fig.1. In the following, we will restrict to the situation with $J_{1}<0$ and $J_{3}>0$, since the most interesting "grey region" exists only in the second quadrant of the $J_{1}-J_{3}$ plane. We will also set $|J_{1}|=1$ as the unit of energy.

\begin{figure}[h!]
\includegraphics[width=7cm,angle=0]{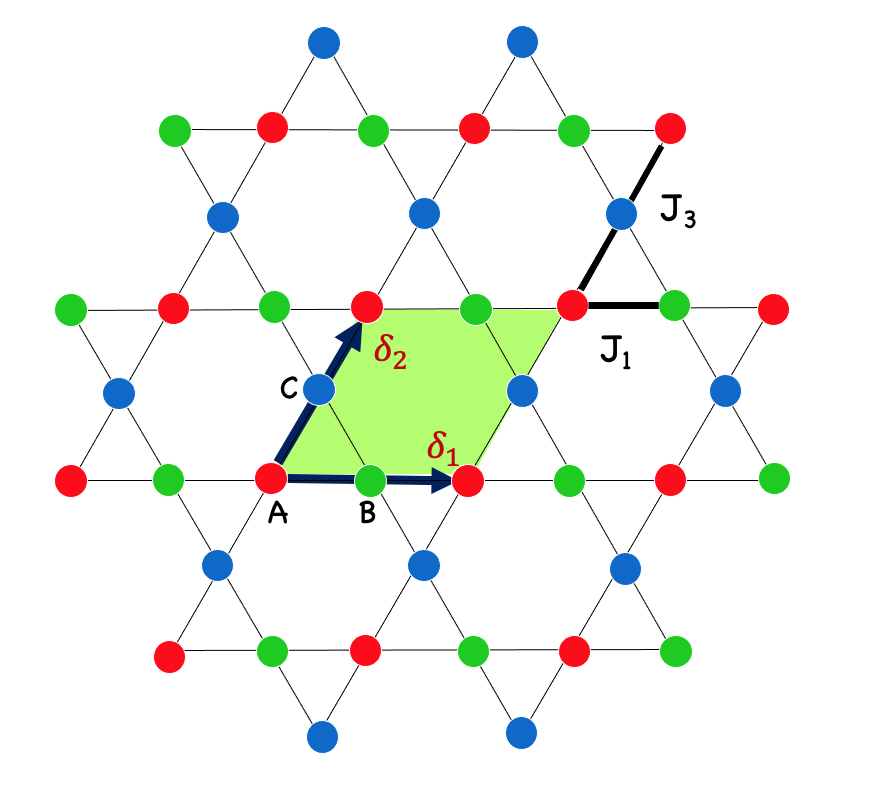}
\caption{The Kagome lattice and the exchange couplings between the first-neighboring and the third-neighboring sites, denoted here as $J_{1}$ and $J_{3}$ respectively. $\bm{\delta}_{1}$, $\bm{\delta}_{2}$ are the two lattice vectors of the Kagome lattice, $\bm{\delta}_{3}=\bm{\delta}_{2}-\bm{\delta}_{1}$. The light green area denotes a unit cell of the Kagome lattice. The lattice sites colored in red, green and blue belong to the three sublattices(namely, sublattice A, B, and C) of the Kagome lattice. In the 3sub-AF phase each of these three sublattices will establish a conventional Neel order as for the SAFHM and the effect of the first-neighboring exchange coupling is exactly cancelled at the classical level. } 
\label{fig1}
\end{figure}

The Fourier transform of the exchange coupling of the $J_{1}-J_{3}$ KAFHM is given by
\begin{equation}
J(\mathbf{q})=2\left(\begin{array}{ccc} 2J_{3}(c^{2}_{1}+c^{2}_{2}) &J_{1}c_{1}&  J_{1}c_{2}\\ J_{1}c_{1}&2J_{3}(c^{2}_{1}+c^{2}_{2})   &J_{1}c_{3} \\ J_{1}c_{2}&J_{1}c_{3}&2J_{3}(c^{2}_{1}+c^{2}_{2})  \end{array}\right)-4J_{3} ,
\end{equation}
in which $c_{i}=\cos \frac{q_{i}}{2}$ and $q_{i}=\mathbf{q}\cdot \bm{\delta}_{i}$. Here, $\bm{\delta}_{1}$, $\bm{\delta}_{2}$ denote the two lattice vectors of the Kagome lattice, $\bm{\delta}_{3}=\bm{\delta}_{2}-\bm{\delta}_{1}$. For $J_{3}\leq \frac{1}{4}$, the lowest eigenvalue of $J(\mathbf{q})$ has the unique minimum at $\mathbf{q}=0$, corresponding to a ferromagnetic ground state of the system. Consistent with this ground state, Monte Carlo simulation find no finite temperature transition in this region\cite{Grison}. For  $J_{3}\geq \frac{1+\sqrt{5}}{4}\approx0.809$, the lowest eigenvalue of $J(\mathbf{q})$ reaches its minimum at three degenerate momentum $\mathbf{q}_{1}=(\pi,0)$, $\mathbf{q}_{2}=(0,\pi)$  and $\mathbf{q}_{3}=(\pi,\pi)$. These three momentums actually correspond to the antiferromagnetic ordering wave vector on the three sublattices(sublattice B, C, and A respectively) of the Kagome lattice. In such a 3sub-AF state, the effect of $J_{1}$ is exactly cancelled at the classical level, no matter what is its sign. The Neel order on the three sublattices can thus rotate freely from each other without any energy penalty. This is alike the situation in the $J_{1}-J_{2}$ SAFHM when $J_{2}\geq\frac{J_{1}}{2}$, where the Neel order of the two antiferromagetic sublattices can rotate freely from each other without any energy penalty. Such a degeneracy can be lifted by the order-by-disorder mechanism in the $J_{1}-J_{3}$ KAFHM, just as what happens in the $J_{1}-J_{2}$ SAFHM. The collinear spin state then selected will spontaneously break the point group symmetry. In our case, such a symmetry breaking is characterized by a $q=4$ Potts order parameter\cite{Grison}. Previous Monte Carlo simulation indeed found a finite temperature transition to a phase with nonzero $q=4$ Potts order in this region\cite{Grison}.

The situation becomes much more tricky in the intermediate region $\frac{1}{4} \leq J_{3} \leq \frac{1+\sqrt{5}}{4}$, namely, the so called "grey region". In this region, the Luttinger-Tisza criteria fails to predict the classical ordering pattern of the system. In fact, the lower bound predicted by the Luttinger-Tisza criteria can never be reached and there is no general way to determine the classical ordering pattern of the system in this region. The lowest eigenvalue of $J(\mathbf{q})$ reaches its minimum at multiple incommensurate momentums that move continuously with the value of $J_{3}$ in the "grey region". Nevertheless, Monte Carlo simulation find that a finite temperature transition toward some symmetry breaking phase still exists inside the "grey region". It is found that the region $0.69 \leq J_{3}\leq \frac{1+\sqrt{5}}{4}$ is characterized by the same $q=4$ Potts order as that of the 3sub-AF phase. On the other hand, the nature of the order for $J_{3}\leq 0.69$ is unclear. It is also unclear how this unknown order is transformed into the $q=4$ Potts order around $J_{3}\approx0.69$. 

The frustration of classical ordering pattern in the "grey region" will enhance the possibility of realizing quantum spin liquid phase in the quantum version of the  
$J_{1}-J_{3}$ KAFHM. However, before discussing such a possibility it is important to have a more thorough understanding of the classical phase diagram.

\section{III. The order parameters for the finite temperature transitions in the $J_{1}-J_{3}$ KAFHM}
\subsection{A. General considerations}
The key step to understand the classical phase diagram of the $J_{1}-J_{3}$ KAFHM is to find the appropriate order parameter for the various phases of the system. According to the Mermin-Wagner theorem, spontaneous breaking of continuous spin rotational symmetry is prohibited at any finite temperature in two dimensional systems. Thus for the $J_{1}-J_{3}$ KAFHM, a finite temperature transition can only occur through the breaking of a discrete symmetry. To describe such symmetry breaking orders, one should investigate the long range correlation of some spin rotationally invariant object, rather than that of spin variable directly.

The discrete symmetry of the $J_{1}-J_{3}$ KAFHM that can be broken at finite temperature includes the lattice translation symmetry, lattice rotation or inversion symmetry and the time reversal symmetry, or the combinations of them. To detect translation symmetry breaking in the spin rotational invariant channel, one can study the long range correlation between local dimer degree of freedoms defined by 
\begin{equation}
C_{\alpha,\beta}(i,j)=\langle D_{i,\alpha}D_{j,\beta}\rangle-\langle D_{i,\alpha}\rangle \langle D_{j,\beta}\rangle,
\end{equation}
in which $D_{i,\alpha}=\mathbf{S}_{i}\cdot\mathbf{S}_{i+\delta_{\alpha}}$, $\delta_{\alpha}$ and $\delta_{\beta}$ are lattice vectors connecting first-neighboring sites on the lattice\cite{dimer}. Such dimer correlations, when appropriately combined, can also be used to detect rotation symmetry breaking. For example, the stripy order of the $J_{1}-J_{2}$ SAFHM with $J_{2}\geq\frac{J_{1}}{2}$ can be detected by calculating the correlation of the nematicity in local spin correlation, namely, 
\begin{eqnarray}
N(i,j)&=&\langle \ (D_{i,x}-D_{i,y})\ (D_{j,x}-D_{j,y})\ \rangle\nonumber\\
&=&C_{x,x}(i,j)-C_{x,y}(i,j)-C_{y,x}(i,j)+C_{y,y}(i,j).\nonumber\\
\end{eqnarray}
When the time reversal symmetry is spontaneously broken, one can study the correlation of the scalar spin chirality defined as 
\begin{equation}
Q_{\alpha,\beta;\alpha',\beta'}(i,j)=\langle \ \bm{\chi}_{i,\delta_{\alpha},\delta_{\beta}}\ \bm{\chi}_{j,\delta_{\alpha'},\delta_{\beta'}}\ \rangle,
\end{equation}
in which $\bm{\chi}_{i,\delta_{\alpha},\delta_{\beta}}=\mathbf{S}_{i}\cdot(\mathbf{S}_{i+\delta_{\alpha}}\times\mathbf{S}_{i+\delta_{\beta}})$ is the scalar spin chirality on the triangle with vertex $i,i+\delta_{\alpha}$ and $i+\delta_{\beta}$. As was found in the previous study\cite{Grison}, our simulation shows that there is no evidence for the spontaneous breaking of the time reversal symmetry in the $J_{1}-J_{3}$ KAFHM. We will thus discard the chiral order parameter and concentrate on the dimer correlation in the following. 

\subsection{B. The $q=4$ Potts order in the 3sub-AF phase and the higher half of the "grey region"}
  
In the 3sub-AF phase, the local dimer is found to exhibit long range correlation simultaneously at the three wave vector $\mathbf{q}_{1}=(0,\pi)$, $\mathbf{q}_{2}=(\pi,0)$ and  $\mathbf{q}_{3}=(\pi,\pi)$, consistent with the existence of the $q=4$ Potts order proposed in Ref.[\onlinecite{Grison}]. More specifically, the antiferromagnetic correlation within each of the three sublattices(denoted as red, green and blue sites in Fig.1) will endow the dimer in the $\bm{\delta}_{1}$, $\bm{\delta}_{2}$ and $\bm{\delta}_{3}$ direction a modulation wave vector of $\mathbf{q}_{1}=(0,\pi)$, $\mathbf{q}_{2}=(\pi,0)$ and  $\mathbf{q}_{3}=(\pi,\pi)$ respectively. A $q=4$ Potts order parameter $\bm{\Sigma}$ can be defined on each elementary triangle of the Kagome lattice with such a staggered factor taken into account\cite{Grison}. On the up-triangle of the $\bm{\mu}$-th unit cell, it is given by
\begin{equation}
\bm{\Sigma}^{\triangle}_{\bm{\mu}}=\left(\begin{array}{c}(-1)^{\mu_{2}}\  \  \ \mathbf{S}_{\bm{\mu},A}\cdot\mathbf{S}_{\bm{\mu},B}
 \\(-1)^{\mu_{1}}\ \ \ \mathbf{S}_{\bm{\mu},A}\cdot\mathbf{S}_{\bm{\mu},C} \\(-1)^{\mu_{1}+\mu_{2}}\ \ \ \mathbf{S}_{\bm{\mu},B}\cdot\mathbf{S}_{\bm{\mu},C}\end{array}\right). 
\end{equation}
Here $\bm{\mu}=(\mu_{1},\mu_{2})$, with $\mu_{1}$ and $\mu_{2}$  the unit cell index in the $\bm{\delta}_{1}$ and $\bm{\delta}_{2}$ direction. Similarly, the $q=4$ Potts order parameter on the down-triangle of the $\bm{\mu}$-th unit cell is given by
\begin{equation}
\bm{\Sigma}^{\bigtriangledown}_{\mu}=\left(\begin{array}{c}(-1)^{\mu_{2}}\  \  \ \mathbf{S}_{\bm{\mu}+(1,1),A}\cdot\mathbf{S}_{\bm{\mu}+(0,1),B}
 \\(-1)^{\mu_{1}}\ \ \ \mathbf{S}_{\bm{\mu}+(1,1),A}\cdot\mathbf{S}_{\bm{\mu}+(1,0),C} \\(-1)^{\mu_{1}+\mu_{2}}\ \ \ \mathbf{S}_{\bm{\mu}+(0,1),B}\cdot\mathbf{S}_{\bm{\mu}+(1,0),C}\end{array}\right). 
\end{equation}
To detect the $q=4$ Potts order, we can calculate the structure factor of $\bm{\Sigma}^{\triangle}$ and $\bm{\Sigma}^{\bigtriangledown}$ defined as follows
\begin{equation}
\bm{\Phi}_{4}=|\frac{1}{2N}\sum_{\bm{\mu}} (\bm{\Sigma}^{\triangle}_{\bm{\mu}}+\bm{\Sigma}^{\bigtriangledown}_{\bm{\mu}})|^{2}.
\end{equation}
Here $N$ denotes the number of unit cells in the system. In the absence of the $q=4$ Potts order we expect $\bm{\Phi}_{4}$ to decrease with $N$ as $\frac{1}{N}$ in the large $N$ limit, while in the ordered phase we expect it to approach a constant value of order one. In particular, if the system choose to order in an perfect collinear pattern in the 3sub-AF phase we expect $\bm{\Phi}_{4}=3$.

\subsection{C. The $q=3$ Potts order in the lower half of the "grey region"}

\begin{figure}[h!]
\includegraphics[width=7cm,angle=0]{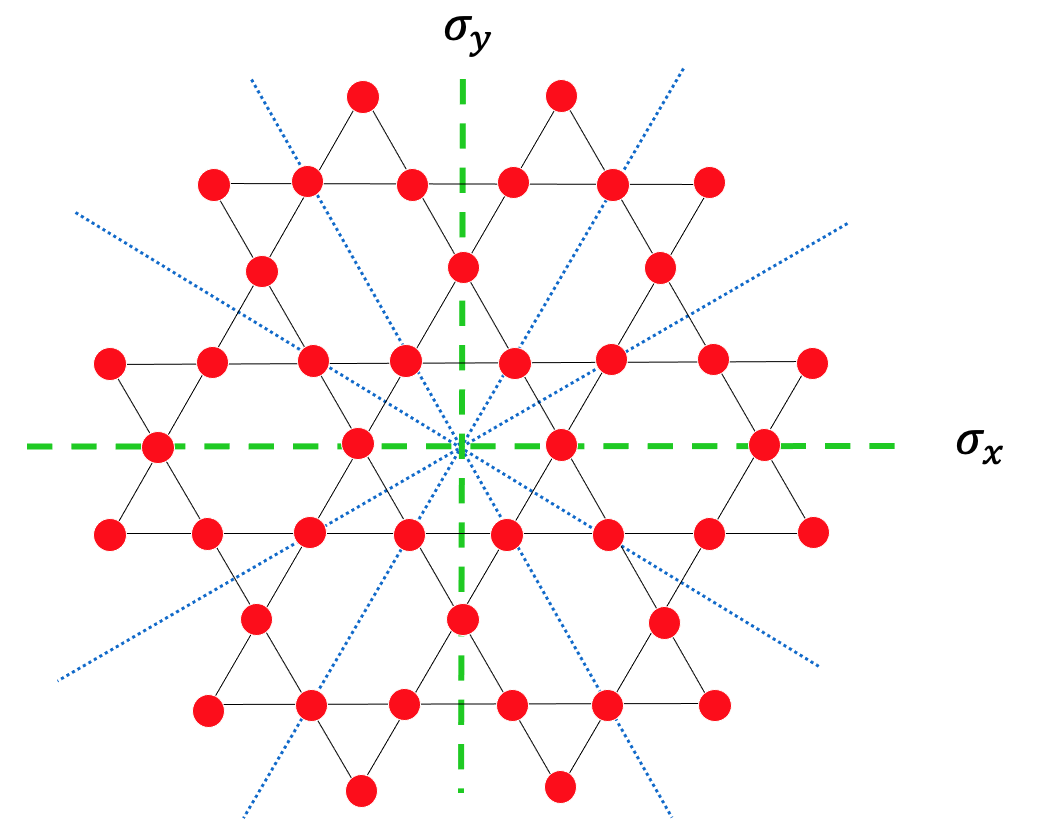}
\caption{The point group symmetry of the $J_{1}-J_{3}$ KAFHM is spontaneously broken from $6mm$ down to $2mm$ in the $q=3$ Potts phase, leaving the dimer in the $\bm{\delta}_{1}$, $\bm{\delta}_{2}$ and $\bm{\delta}_{3}$ direction not all equivalent. Here the long dashed green lines denote the reflection lines of the $2mm$ group. Together with the short dashed blue lines they form the reflection lines of the $6mm$ group.} 
\label{fig2}
\end{figure}

Different from the 3sub-AF phase, we find that the local dimer exhibits long range correlation only at $\mathbf{q}=0$ in the lower half of the "grey region". As will be detailed in the next section, we find that the point group symmetry of the $J_{1}-J_{3}$ KAFHM is spontaneously broken from $6mm$ down to $2mm$, leaving the dimer in the $\bm{\delta}_{1}$, $\bm{\delta}_{2}$ and $\bm{\delta}_{3}$ direction not all equivalent. A $q=3$ Potts order will thus emerge, which can be defined on the up-triangle as
\begin{eqnarray}
\bm{\psi}^{\triangle}_{\bm{\mu}}&=&\ \ \ \ \ \ \ \ \ \mathbf{S}_{\bm{\mu},A}\cdot\mathbf{S}_{\bm{\mu},B}\nonumber\\
&+&e^{i\frac{2\pi}{3}} \ \ \ \ \mathbf{S}_{\bm{\mu},A}\cdot\mathbf{S}_{\bm{\mu},C}\nonumber\\
&+&e^{-i\frac{2\pi}{3}}\ \ \ \mathbf{S}_{\bm{\mu},B}\cdot\mathbf{S}_{\bm{\mu},C},
\end{eqnarray} 
Different from $\bm{\Sigma}$, which is a vector, here $\bm{\psi}$ is a complex number. The $q=3$ Potts order on the down-triangle can be defined as
\begin{eqnarray}
\bm{\psi}^{\bigtriangledown}_{\bm{\mu}}&=&\ \ \ \ \ \ \ \ \mathbf{S}_{\bm{\mu}+(1,1),A}\cdot\mathbf{S}_{\bm{\mu}+(0,1),B}\nonumber\\
&+&e^{i\frac{2\pi}{3}} \ \ \mathbf{S}_{\bm{\mu}+(1,1),A}\cdot\mathbf{S}_{\bm{\mu}+(1,0),C}\nonumber\\
&+&e^{-i\frac{2\pi}{3}}\ \mathbf{S}_{\bm{\mu}+(0,1),B}\cdot\mathbf{S}_{\bm{\mu}+(1,0),C}.
\end{eqnarray} 
To detect such a $q=3$ Potts order, we can calculate the structure factor of $\bm{\psi}^{\triangle}_{\bm{\mu}}$ and $\bm{\psi}^{\bigtriangledown}_{\bm{\mu}}$ defined as follows
\begin{equation}
\bm{\Phi}_{3}=|\frac{1}{2N}\sum_{\bm{\mu}}({\psi}^{\triangle}_{\bm{\mu}}+\bm{\psi}^{\bigtriangledown}_{\bm{\mu}})|^{2}
\end{equation}
 In the absence of the $q=3$ Potts order we expect $\bm{\Phi}_{3}$ to decrease with $N$ as $\frac{1}{N}$ in the large $N$ limit, while in the ordered phase we expect it to approach a constant value of order one. 
 
\section{IV. The full classical phase diagram of the $J_{1}-J_{3}$ KAFHM}

\subsection{A. The heat bath algorithm and the over relaxation technique}
The classical $J_{1}-J_{3}$ KAFHM is simulated with the Monte Carlo method. In our simulation, we have adopted the heat bath algorithm combined with the over relaxation technique\cite{Young,Okubo}. We find that such a choice is particularly convenient for the simulation of the $J_{1}-J_{3}$ KAFHM. It performs much better than other local update algorithm that we have attempted. We note that such an algorithm has been successfully applied to simulate very large system\cite{Okubo}. 

In each heat bath update step, we draw randomly a lattice site $i$ and calculate the molecular field $\mathbf{H}_{i}$ acting on it from the neighboring sites through the exchange coupling, which is given by
\begin{equation}
\mathbf{H}_{i}=\frac{1}{T}\sum_{j}J_{i,j}\mathbf{S}_{j}.
\end{equation}
$\mathbf{S}_{i}$ is then updated to a new direction according to the local thermal distribution under the action of $\mathbf{H}_{i}$, namely
\begin{equation}
P(\theta,\phi)=\frac{1}{4\pi}\frac{|\mathbf{H}_{i}|}{\sinh |\mathbf{H}_{i}|}e^{|\mathbf{H}_{i}|\cos\theta},
\end{equation}
in which $\theta$ and $\phi$ are the polar and azimuthal angle of $\mathbf{S}_{i}$ relative to $\mathbf{H}_{i}$. Such a distribution can be achieved by choosing $\theta$ and $\phi$ as follows\cite{Young}
\begin{eqnarray}
\cos\theta&=&\frac{1}{|\mathbf{H}_{i}|} \mathrm{Ln} [1+R_{1}(e^{|\mathbf{H}_{i}|}-1)]-1\nonumber\\
\phi&=&2\pi R_{2},
\end{eqnarray}
in which $R_{1}$ and $R_{2}$ are two random number uniformly distributed between 0 and 1. The advantage of the heat bath algorithm over the conventional Metropolis algorithm lies in the fact that each heat bath update is accepted with probability 1.

In between each step of heat bath update, we can insert a number of over relaxation update steps with extremely low computational cost. In each over relaxation update step, we choose randomly a lattice site $i$ and reverse the perpendicular component of $\mathbf{S}_{i}$ relative to the local molecular field $\mathbf{H}_{i}$. Since there is no change in the energy, the updated spin is accepted with probability 1. In realistic simulation, we have inserted 9 over relaxation steps between each pair of heat bath update step.

\subsection{B. The classical phase diagram of the $J_{1}-J_{3}$ KAFHM}
The classical phase diagram of the  $J_{1}-J_{3}$ KAFHM can be mapped out from the calculation of the specific heat of the system, which is given by
\begin{equation}
C_{V}=\frac{1}{NT^{2}}(\langle E^{2} \rangle-\langle E \rangle^{2}),
\end{equation}
in which $E$ is the local energy of a spin configuration. The phase boundary in the phase diagram manifests itself in the form of a sharp peak at a finite temperature transition point. For a first order transition, the specific heat peak will evolve into a $\delta$-function peak with nonzero integrated area in the thermodynamic limit, which corresponds to the latent heat of the phase transition. Correspondingly, the internal energy will exhibit a finite jump across the transition point. However, on a finite system it is impossible to distinguish a broadened $\delta$-function peak of a first order transition from the smeared specific heat peak of a conventional second order transition. In such a case, evidence for the first order nature of a phase transition can instead be found in the bimodal distribution in the histogram of local energy\cite{Grison}.
 
 \begin{figure}[h!]
\includegraphics[width=7cm,angle=0]{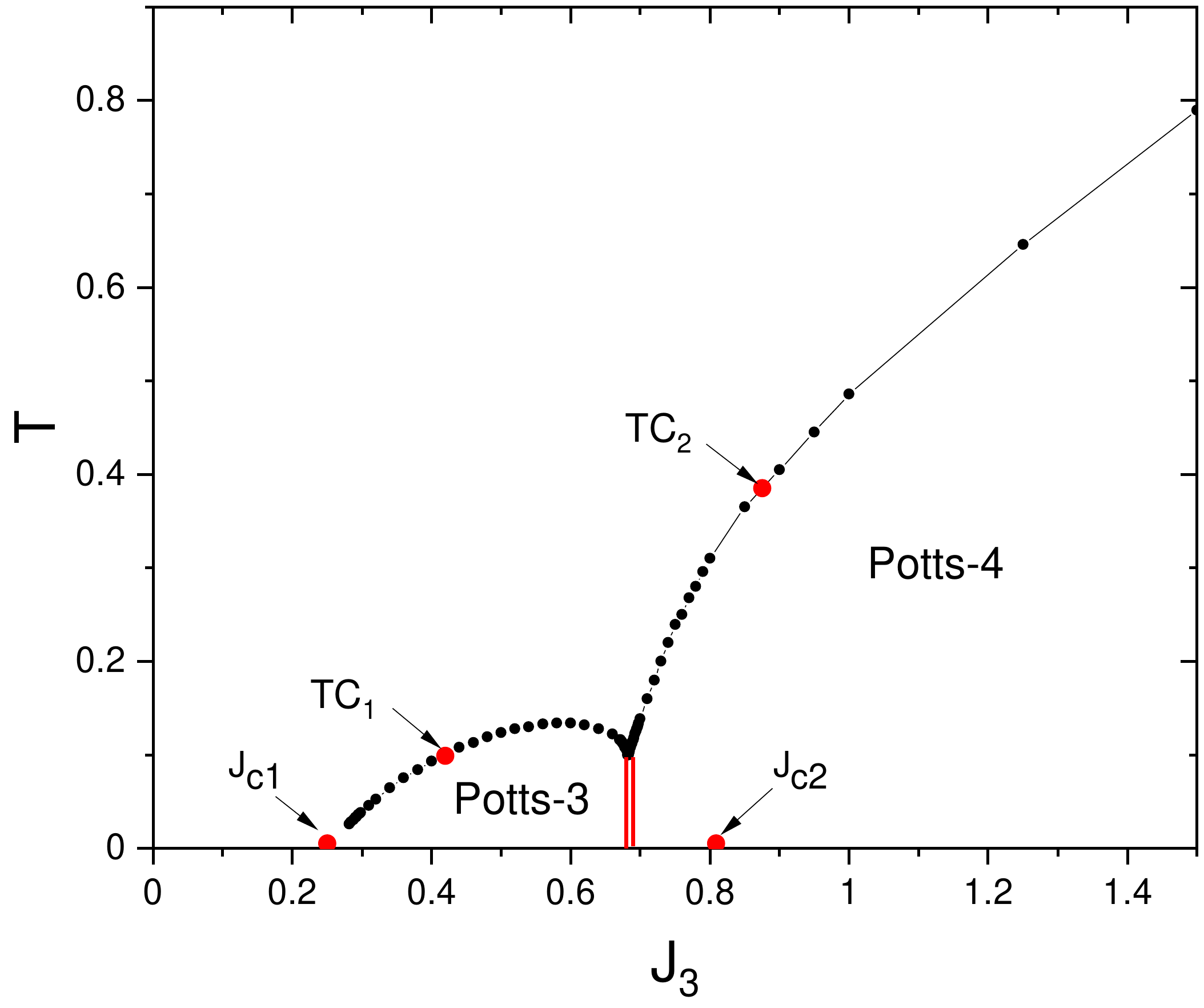}
\caption{The classical phase diagram of the $J_{1}-J_{3}$ KAFHM determined from Monte Carlo simulation. Here $J_{1}=-1$ is set as the unit of energy. $J_{c1}=\frac{1}{4}$ and $J_{c2}=\frac{1+\sqrt{5}}{4}$ are the lower and the upper boundary of the "grey region" in which the Luttinger-Tisza criteria fails to predict the classical ground state of the model. TC$_{1}$ and TC$_{2}$ are two tricritical point along the phase boundary. The phase transition between TC$_{1}$ and TC$_{2}$ is first order, as evidenced by the bimodal distribution of local energy. The phase transition below TC$_{1}$ and that above TC$_{2}$ are found to be continuous. The doubled red line at $J_{3}\approx 0.69$ denotes a first order transition between a $q=3$ Potts phase that break the three-fold rotation symmetry but remain translational invariant and a $q=4$ Potts phase  that break the translational symmetry and the $K_{4}$ symmetry as defined in Ref.[\onlinecite{Grison}].} 
\label{fig3}
\end{figure}
 
The classical phase diagram of the $J_{1}-J_{3}$ KAFHM is presented in Fig.3. The simulations are done on $L\times L\times 3$ lattice with $L=24,48,96,192$. In most Monte Carlo runs, we have used $2\times10^{9}$ local update steps to thermalize the system. The expectation values are calculated using $4.8\times10^{6}$ samples, with each sample drawn from $10^{4}$ local update steps. To keep good statistics in the result, our simulation is restricted to $T\geq 0.03$.

There are three phases in the classical phase diagram, namely the symmetric phase emanating from the ferromagnetic ground state for $J_{3}\leq J_{c1}=\frac{1}{4}$, the Potts-3 phase that break the three-fold rotation symmetry but remain translational invariant and the Potts-4 phase that break the translational symmetry and the $K_{4}$ symmetry as defined in Ref.[\onlinecite{Grison}]. The Potts-3 and the Potts-4 phase are connected by an almost vertical first order phase transition line at $J_{3}\approx0.69$. From the phase diagram it is clear that the Potts-4 phase is not restricted to $J_{3}\geq J_{c2}=\frac{1+\sqrt{5}}{4}$ and can penetrate substantially into the "grey region". Similarly, the ferromagnetic ground state may survive slightly beyond the Luttinger-Tisza boundary $J_{c1}=\frac{1}{4}$, although an accurate determination of the transition point is challenging.

 \begin{figure}[h!]
\includegraphics[width=8cm,angle=0]{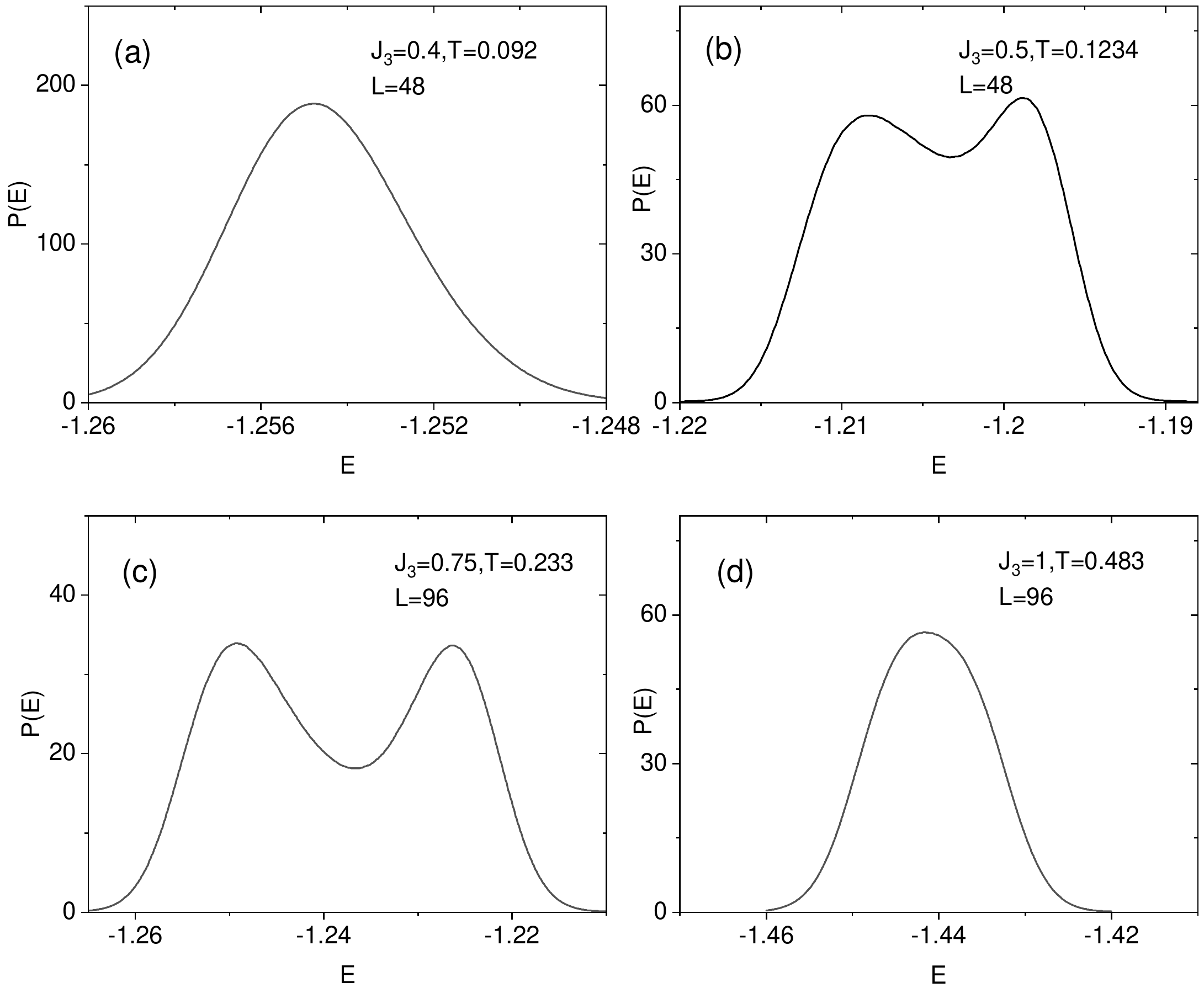}
\caption{The distribution of local energy at the transition temperature for $J_{3}=0.4$, $0.5$, $0.75$ and $1$. The bimodal distribution in the local energy at $J_{3}=0.5$ and $J_{3}=0.75$ is the clear signature of a first order transition at these two points.} 
\label{fig4}
\end{figure}

Along the phase boundary we find two tricritical points at TC$_{1}$ and TC$_{2}$.  TC$_{1}\approx0.42$ is located in the Potts-3 region and TC$_{2}\approx0.87$ is located in the Potts-4 region. The phase transition line in between TC$_{1}$ and TC$_{2}$ is found to be first order, as evidenced by the bimodal distribution of the local energy at the transition point. The phase transition line below TC$_{1}$ and that above TC$_{2}$ are found to be continuous. The histogram of local energy at the transition point is plotted in Fig.4 at four representative points in the phase diagram, namely at $J_{3}=0.4$, $0.5$, $0.75$ and $1$. The mechanism through which such tricritical behavior emerge is now unknown to us.

Different from the speculation made in Ref.[\onlinecite{Grison}], we find no reentrance behavior for the Potts-4 order between $J_{3}\approx 0.69$ and $J_{c2}=\frac{1+\sqrt{5}}{4}$. The phase boundary between the Potts-3 phase and the Potts-4 phase is found to be first order and almost independent of temperature. The upper boundary of the "grey region" predicted by the Luttinger-Tisza analysis plays no role in the phase diagram.

\subsection{C. The structure of the symmetric phase}
While a finite temperature phase transition is prohibited by the Mermin-Wanger theorem in this region, it is still interesting to investigate the evolution of the spin structure factor with $J_{3}$ at low temperature. For the $J_{1}-J_{3}$ KAFHM, we can define the spin structure factor of the system as the trace of the spin correlation matrix, or
\begin{equation}
S(\mathbf{q})=\frac{1}{N}\sum_{i,\mu}e^{i\mathbf{q}\cdot \mathbf{R}_{i}}\langle \mathbf{S}_{i,\mu} \cdot \mathbf{S}_{j,\mu}\rangle,
\end{equation}
in which $\mu=A,B,C$ denotes the sublattice index.

 \begin{figure}[h!]
\includegraphics[width=9cm,angle=0]{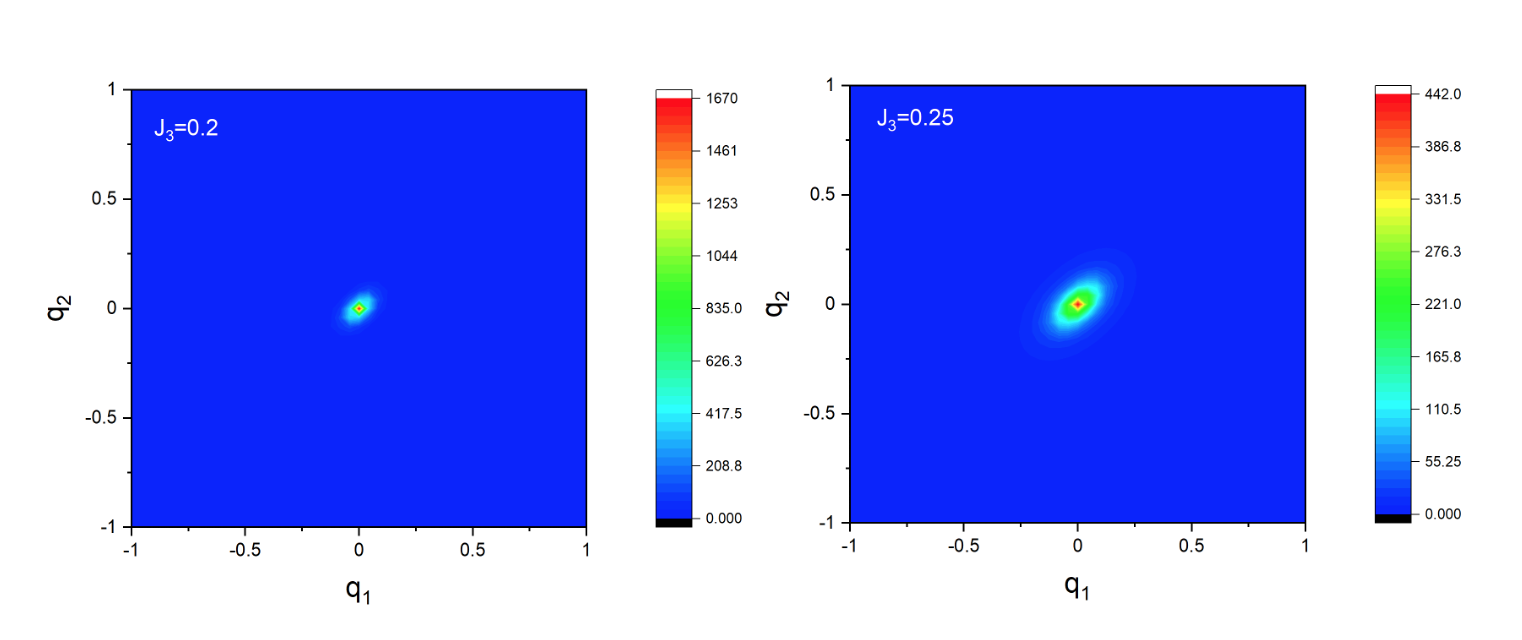}
\includegraphics[width=9cm,angle=0]{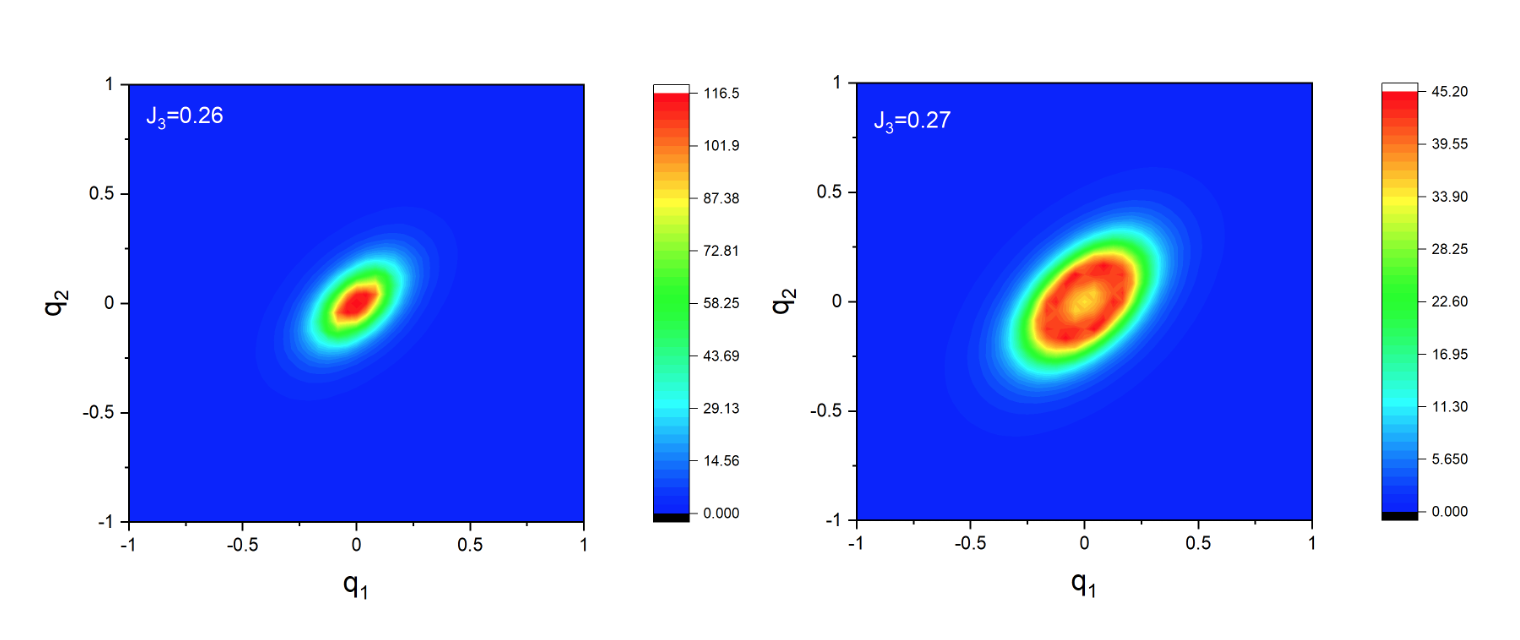}
\includegraphics[width=9cm,angle=0]{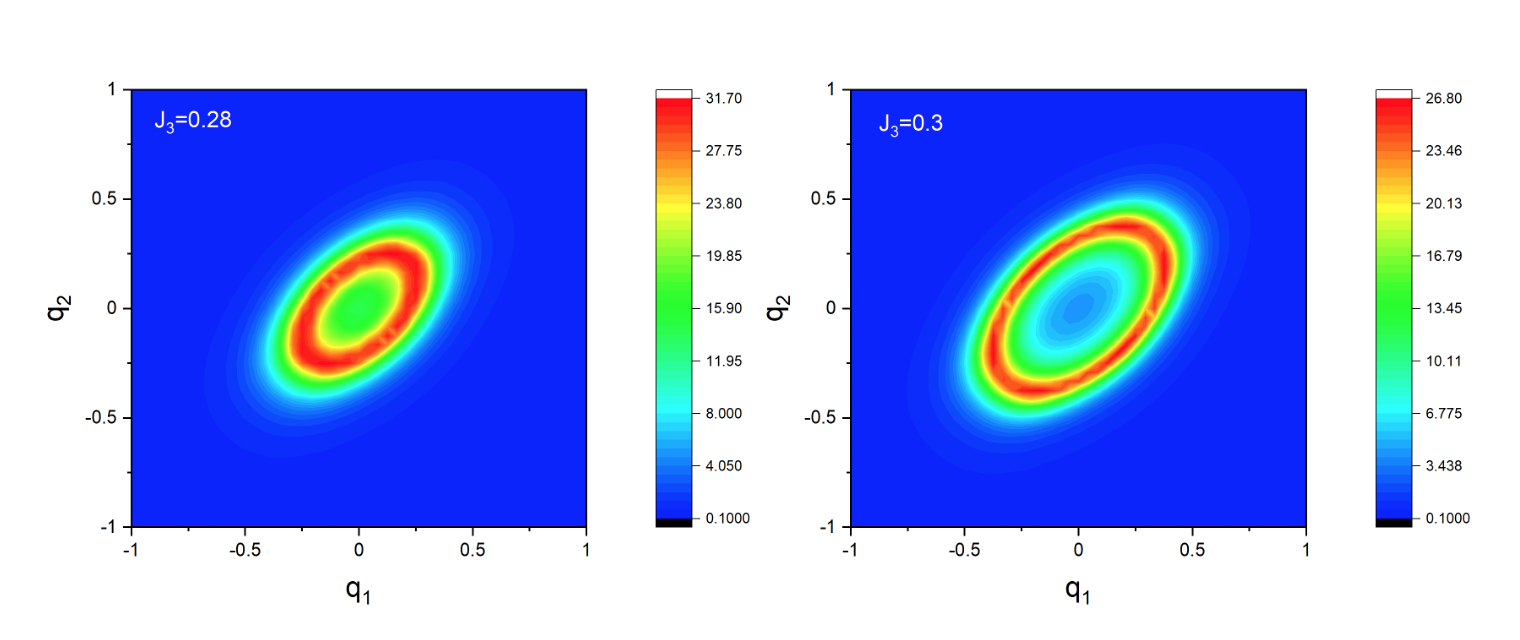}
\caption{The evolution of the spin structure factor in the symmetric phase with $J_{3}$. $T$ is fixed at $0.05$. The calculation is done on a $L=48$ lattice. Note that $T=0.05$ is still above the transition temperature toward the Potts-3 phase for $J_{3}=0.3$. $q_{1}$ and $q_{2}$ are measured in unit of $\pi$.} 
\label{fig5}
\end{figure}

We have computed the spin structure factor at $T=0.05$. The result is shown Fig.5 for a series of $J_{3}$ value within the symmetric phase. For small $J_{3}$, $S(\mathbf{q})$ is found to be essentially indistinguishable from that of the ferromagnetic long range ordered phase, namely, a sharp peak at $\mathbf{q}=0$. Starting from $J_{3}=\frac{1}{4}$, the single peak at $\mathbf{q}=0$ evolves into a ring structure around it. The radius of the ring increases rapidly with the increase of $J_{3}$. Such a ring structure is similar to that observed in the double spiral phase of doped t-J model\cite{Kane}.

\subsection{D. The structure of the Potts-3 phase}

 \begin{figure}[h!]
\includegraphics[width=9cm,angle=0]{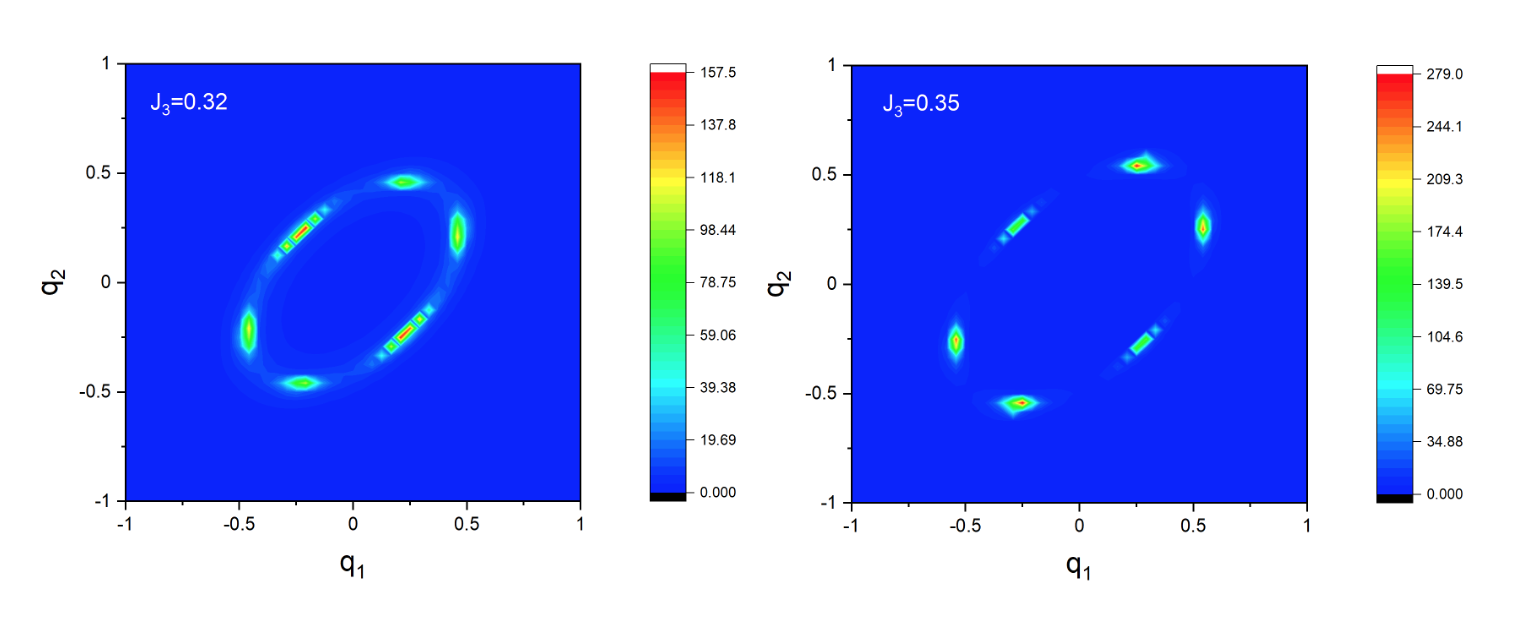}
\includegraphics[width=9cm,angle=0]{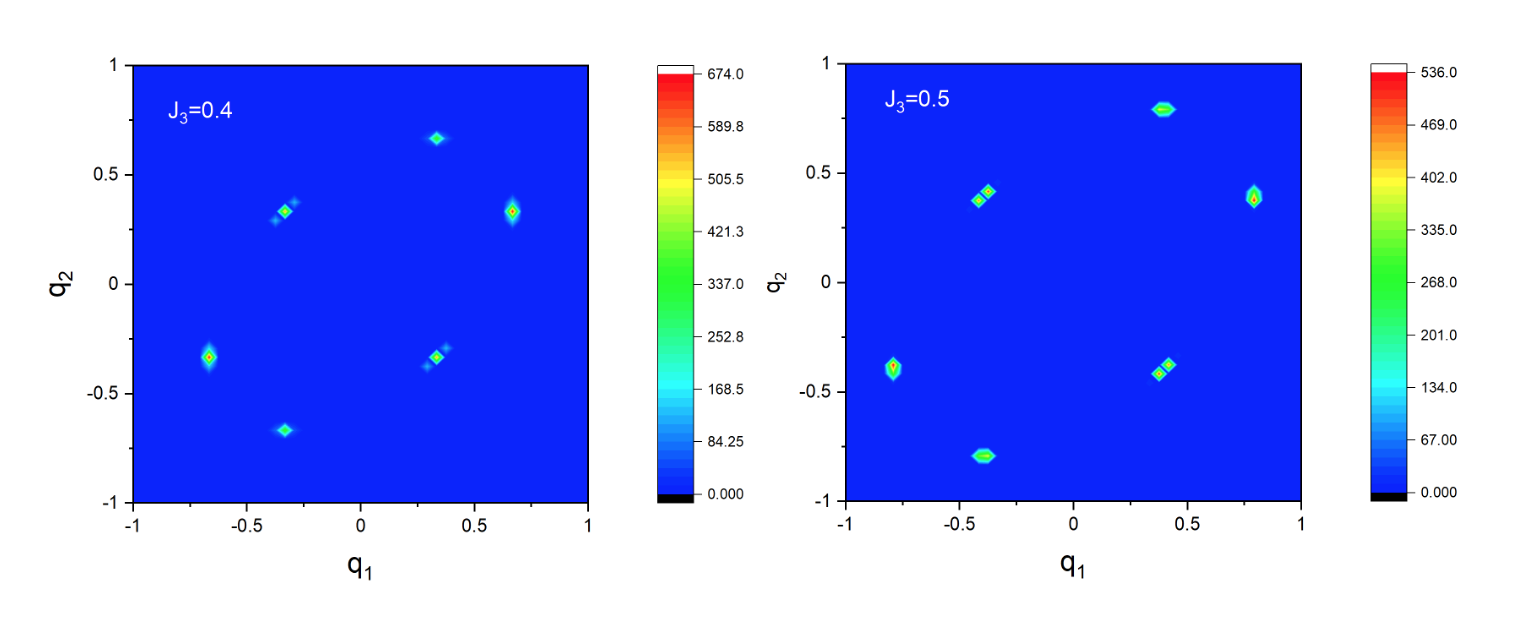}
\includegraphics[width=9cm,angle=0]{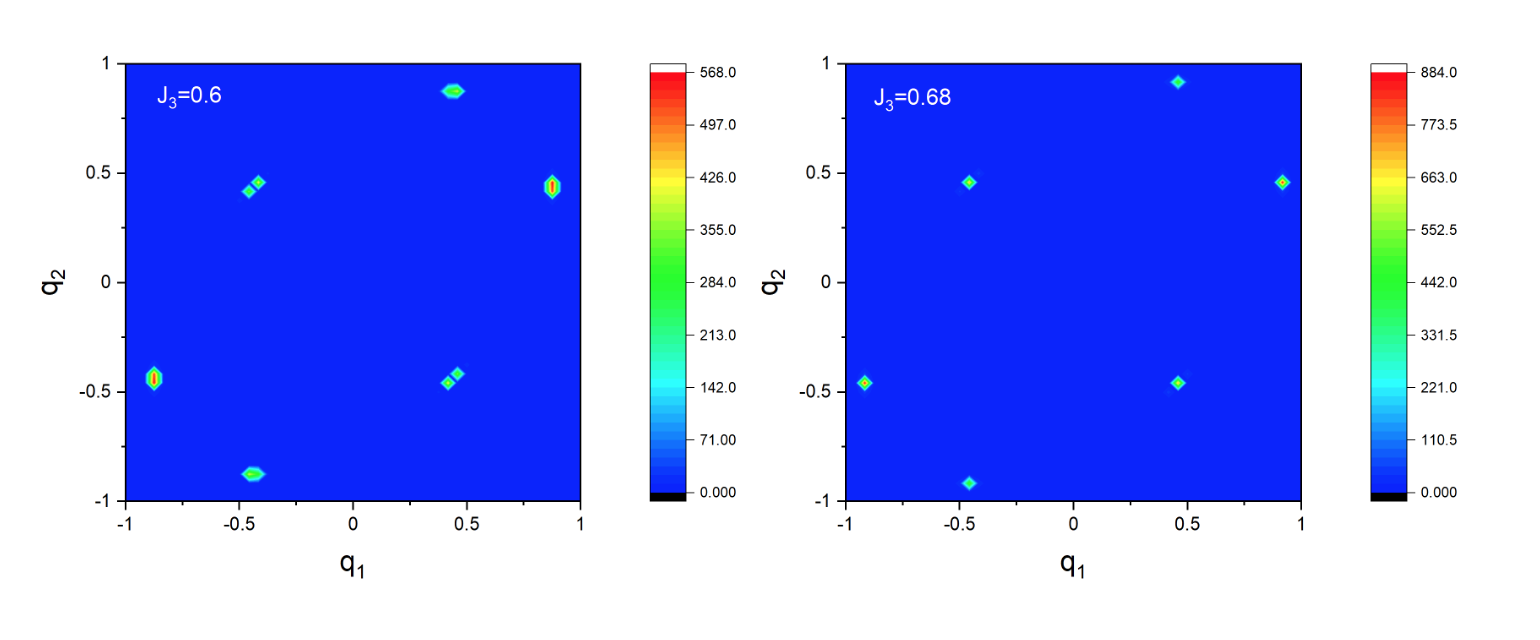}
\caption{The evolution of the spin structure factor in the Potts-3 phase with $J_{3}$. $T$ is fixed at $0.05$. The calculation is done on a $L=48$ lattice. Note that $T=0.05$ is slightly below the transition temperature toward the Potts-3 phase for $J_{3}=0.32$. $q_{1}$ and $q_{2}$ are measured in unit of $\pi$.} 
\label{fig6}
\end{figure}

The evolution of the spin structure factor in the Potts-3 phase is presented in Fig.6. The ring structure observed in Fig.5 now becomes increasingly anisotropic with the increase of $J_{3}$ and eventually evolves into six very sharp peaks. Such anisotropy in the spin structure factor hints at spatial symmetry breaking. However, we note that the sharp peaks in Fig.6 should not be understood as Bragg peak, since spontaneous breaking of the spin rotational symmetry can never happen at finite temperature according to the Mermin-Wagner theorem.

\begin{figure}[h!]
\includegraphics[width=9cm,angle=0]{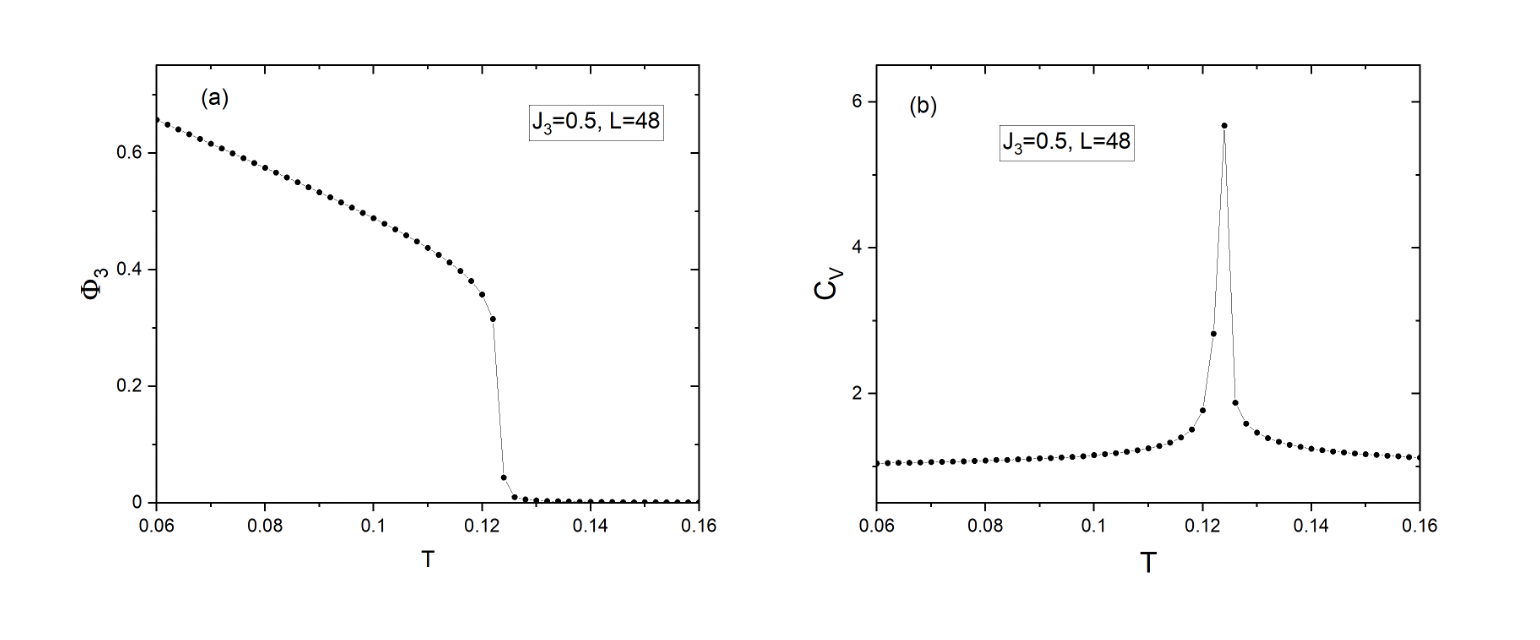}
\includegraphics[width=9cm,angle=0]{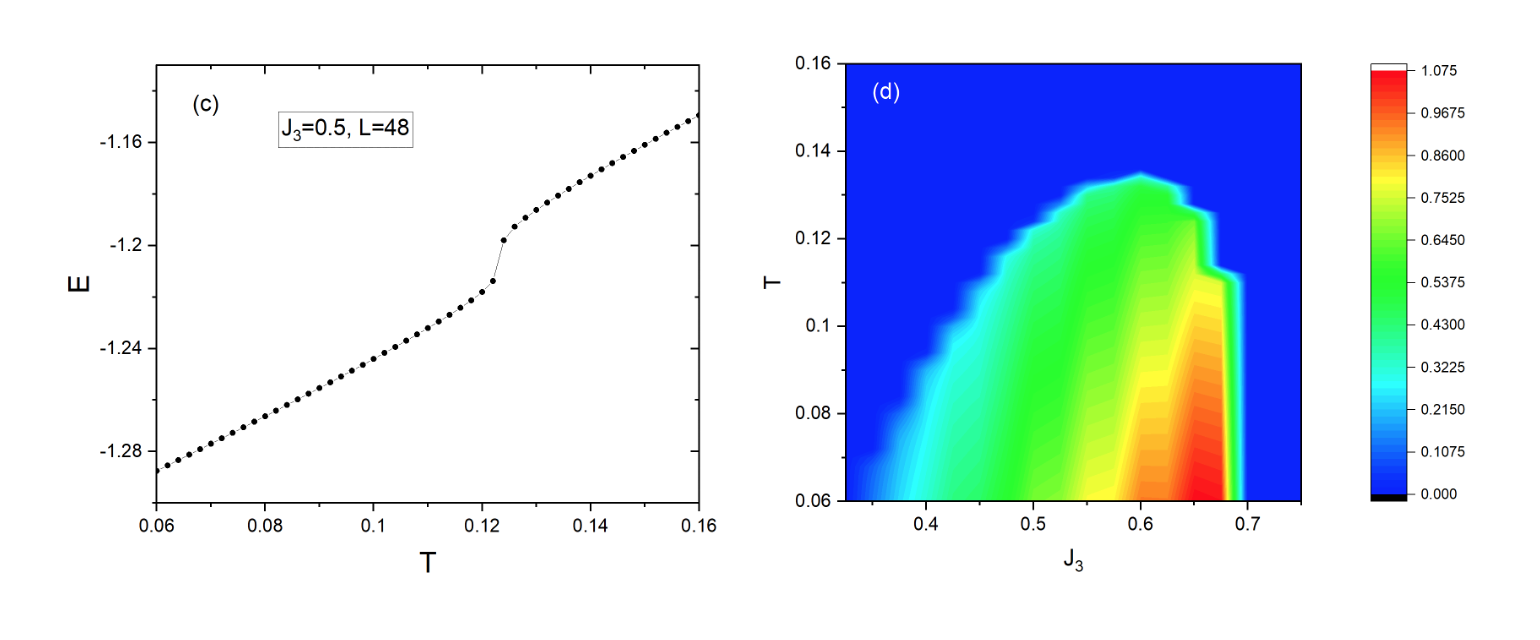}
\caption{The evolution of (a)the Potts-3 order parameter $\bm{\Phi}_{3}$, (b)the specific heat and (c)the internal energy of the $J_{1}-J_{3}$ KAFHM with temperature at $J_{3}=0.5$. The calculation is done on a $L=48$ lattice. In (d) we illustrate the full dependence of the Potts-3 order parameter $\bm{\Phi}_{3}$ on $J_{3}$ and temperature in color scale.} 
\label{fig7}
\end{figure}

To detect the spatial symmetry breaking in the Potts-3 phase, we have calculated the structure factor of the Potts-3 order parameter defined in Sec.IIIC. The result is presented in Fig.7. As can be seen from the plot, the Potts-3 order parameter vanishes abruptly around $J_{3}\approx 0.69$. It is also clear that on the $L=48$ lattice the Potts-3 order parameter is not a smooth function of $J_{3}$. To find the origin of such a non-smoothness behavior, we have compared the Potts-3 order parameter for system with $L=24,48,96$ and $192$, the result of which is shown in Fig.8. We find while the transition temperature is a quite smooth function of $J_{3}$, the non-smoothness in the Potts-3 order parameter is rather significant on finite system. It vanishes only on rather large system. Such a non-smoothness behavior is not at all surprising considering the incommensurability in the Luttinger-Tisza momentum in the "grey region".

\begin{figure}[h!]
\includegraphics[width=9cm,angle=0]{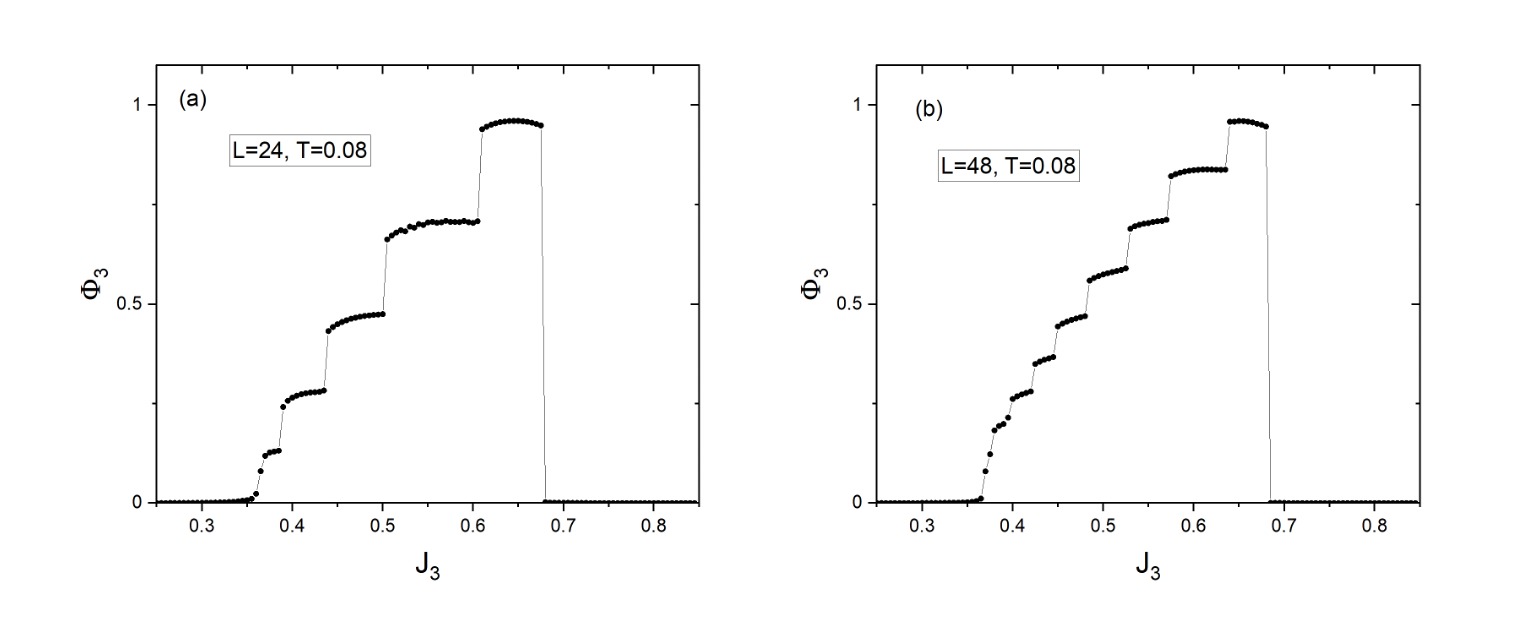}
\includegraphics[width=9cm,angle=0]{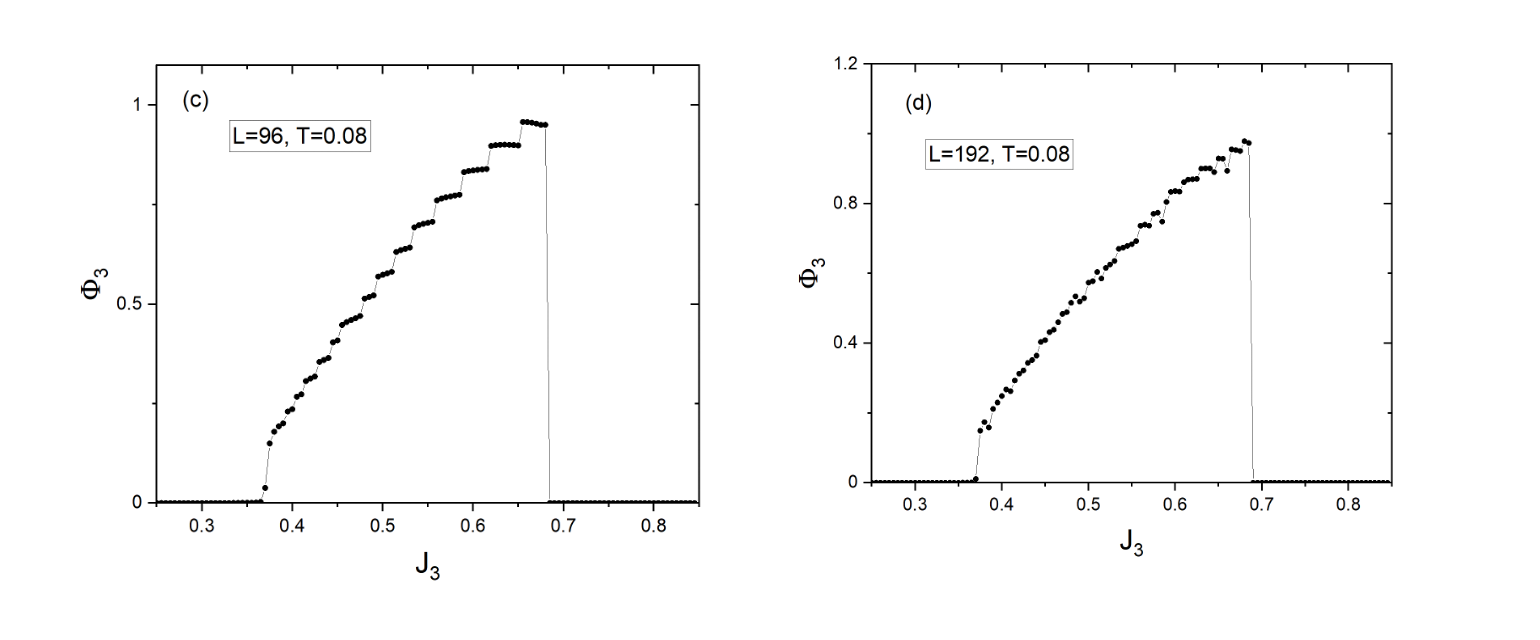}
\caption{The evolution of  Potts-3 order parameter with $J_{3}$ for system with $L=24,48,96$ and $192$. $T$ is fixed at 0.08 in all the calculations. To achieve good statistics in the result, we have used $5\times 10^{10}$ thermalization steps in the calculation of the $L=24,48,96$ system, and used $2\times 10^{11}$ thermalization steps in the calculation of the $L=192$ system. The number of samples used to compute the expectation value is $10^{7}$ for the $L=24,48,96$ system and $4\times 10^{7}$ for the $L=192$ system. Each sample is drawn from $10^{4}$ local updates. } 
\label{fig8}
\end{figure}

 \begin{figure}[h!]
\includegraphics[width=9cm,angle=0]{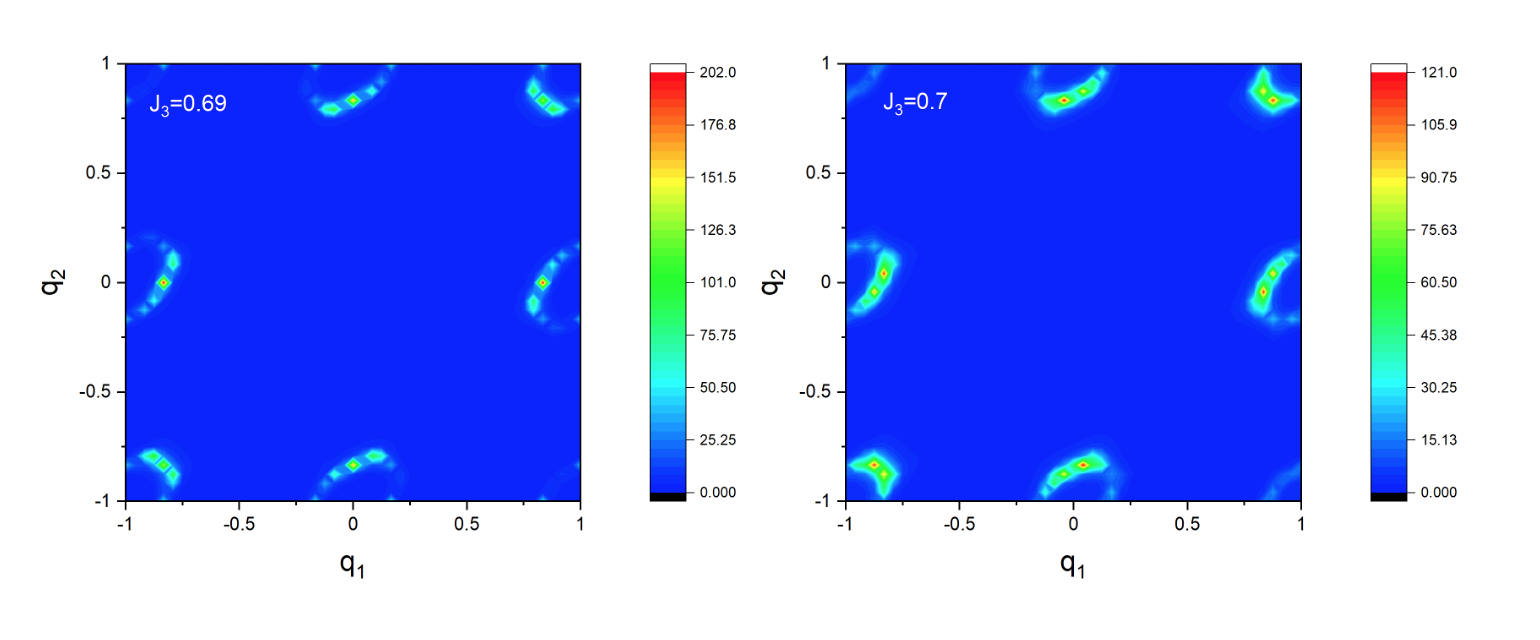}
\includegraphics[width=9cm,angle=0]{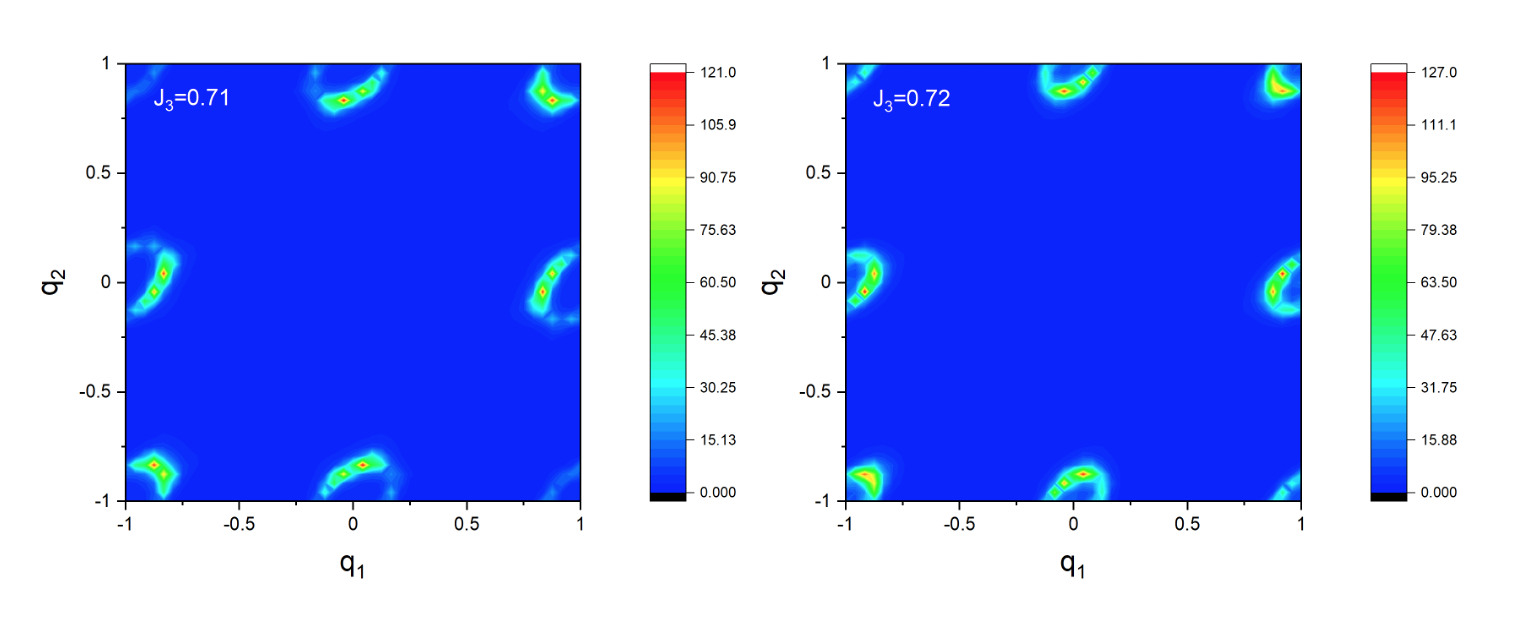}
\includegraphics[width=9cm,angle=0]{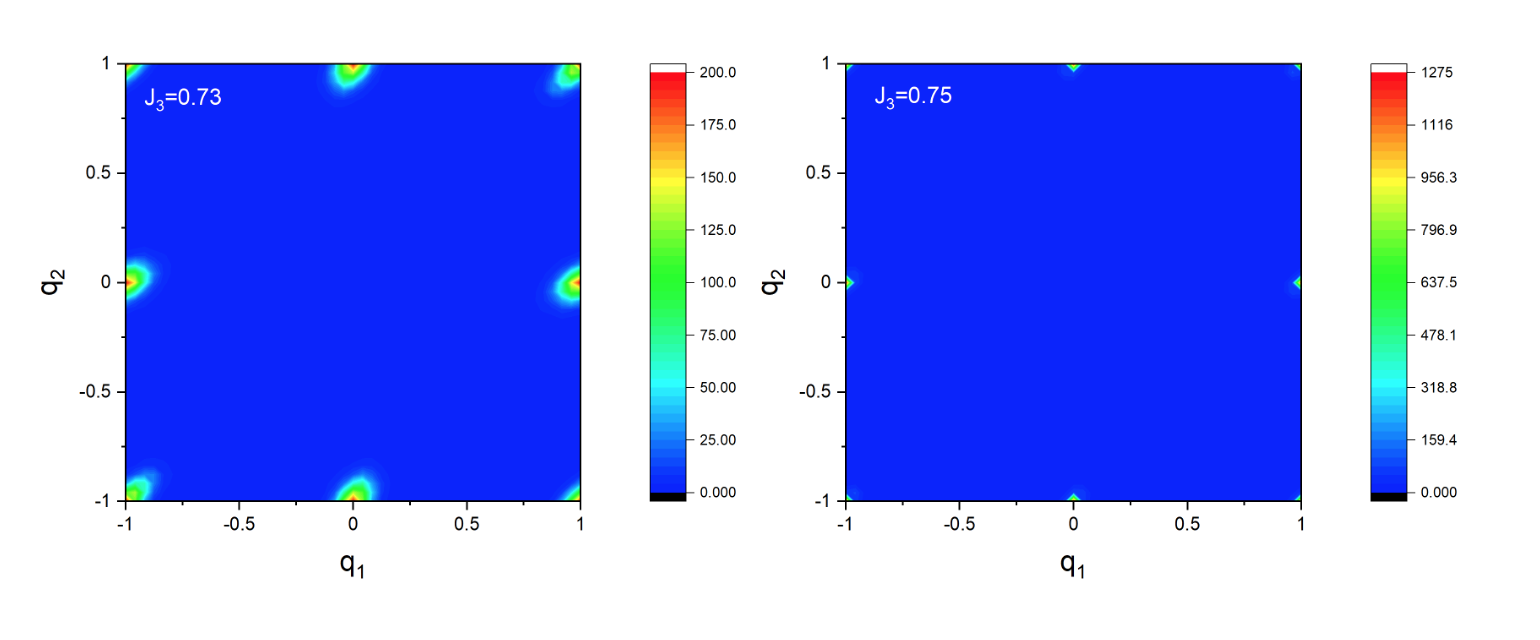}
\caption{The evolution of the spin structure factor in the Potts-4 phase with $J_{3}$. $T$ is fixed at $0.05$. The calculation is done on a $L=48$ lattice.  $q_{1}$ and $q_{2}$ are measured in unit of $\pi$.} 
\label{fig9}
\end{figure}

\subsection{E. The structure of the Potts-4 phase}
The evolution of the spin structure factor in the Potts-4 phase at $T=0.05$ is presented in Fig.9. Just across the phase boundary between the Potts-3 and the Potts-4 phase, the spin stucture factor features three small rings around the momentum $\mathbf{q}_{1}=(\pi,0)$, $\mathbf{q}_{2}=(0,\pi)$ and $\mathbf{q}_{3}=(\pi,\pi)$. With the further increase of $J_{3}$, the radius of these rings rapidly shrinks and for $J_{3}\geq 0.73$ they evolves into three sharp peaks at $\mathbf{q}_{1}=(\pi,0)$, $\mathbf{q}_{2}=(0,\pi)$ and $\mathbf{q}_{3}=(\pi,\pi)$. The spin structure factor almost saturates for $J_{3}\geq0.85$. Such evolution in the spin structure factor implies that the Potts-4 order can exist before a commensurate antiferromagnetic correlation within each of the three sublattices is establishes. 

\begin{figure}[h!]
\includegraphics[width=9cm,angle=0]{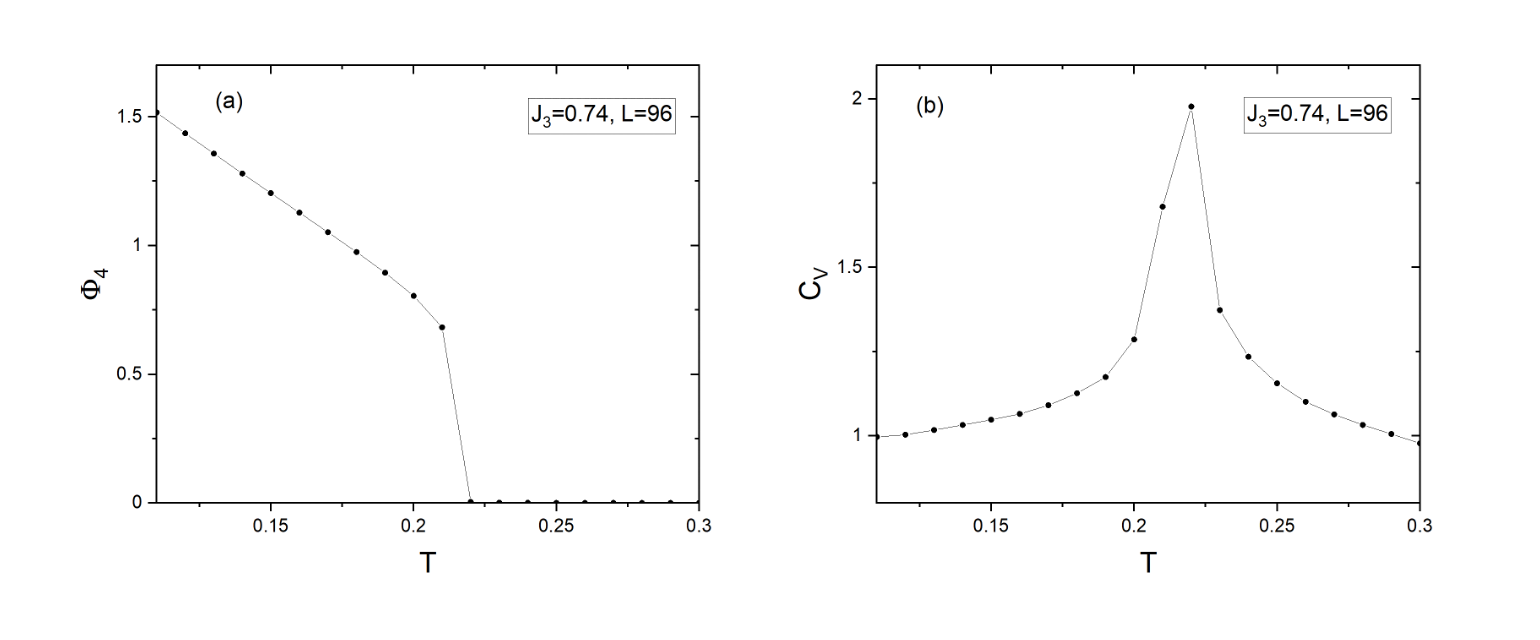}
\includegraphics[width=9cm,angle=0]{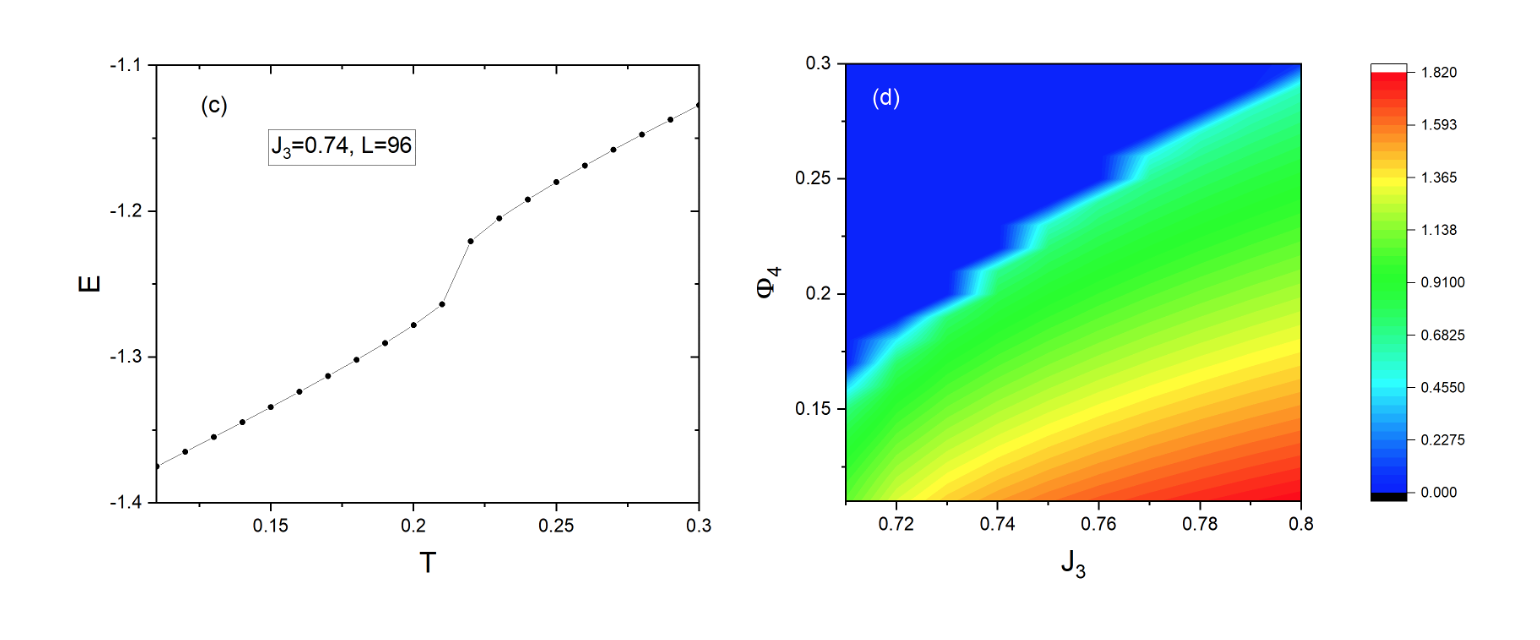}
\caption{The evolution of (a)the Potts-4 order parameter $\bm{\Phi}_{4}$, (b)the specific heat and (c)the internal energy of the $J_{1}-J_{3}$ KAFHM with temperature at $J_{3}=0.74$. The calculation is done on a $L=96$ lattice. In (d) we illustrate the full dependence of the Potts-4 order parameter $\bm{\Phi}_{4}$ on $J_{3}$ and temperature in color scale.} 
\label{fig10}
\end{figure}

The results for the Potts-4 order parameter is presented in Fig.10. Different from the Potts-3 order parameter, we find that the Potts-4 order parameter is a rather smooth function of $J_{3}$ at a fixed temperature.

\subsection{F. Some general comments on the classical phase diagram of the $J_{1}-J_{3}$ KAFHM}
We close the discussion on the classical phase diagram of the $J_{1}-J_{3}$ KAFHM with some general comments. 

First, the Potts-4 order is found to penetrate substantially into the "grey region" with no evidence of a phase transition at the Luttinger-Tisza boundary $J_{c2}=\frac{1+\sqrt{5}}{4}$. The spin structure factor of the system takes the form of small rings in a finite region in the Potts-4 phase. This is quite different from the situation in the $J_{1}-J_{2}$ SAFHM, in which the stripy order emerges only when strong and commensurate antiferromagnetic correlation has already been established within each of its two sublattices. Thus, the emergence of the Potts-4 order in the  $J_{1}-J_{3}$ KAFHM can not be totally attributed to the conventional order-by-disorder mechanism. This is even more obvious for the Potts-3 order, which emerges entirely in the "grey region". In fact, both orders can be thought to be driven by the preference of the system to resolve the strong frustration in the "grey region". 

Second, we note that the spin structure factor of the system is characterized by some anisotropic ring structure in the whole "grey region". In the Potts-3 phase, the ring is found to be centered around $\mathbf{q}=0$. In the Potts-4 phase, the rings are found to be centered around $\mathbf{q}_{1}=(\pi,0)$, $\mathbf{q}_{2}=(0,\pi)$ and $\mathbf{q}_{3}=(\pi,\pi)$. We note that the spin structure factor is calculated at $T=0.05$ rather than at zero temperature. The region where the ring structure is observed($0.27\leq J_{3} \leq 0.72$) is smaller than the "grey region". We guess when the computation is performed at exactly zero temperature these two regions will coincide with each other. Thus, while the Luttinger-Tisza boundary is not a real phase boundary, it determines the behavior of the spin structure factor of the system.

Third, while the evidence for the existence of the two tricritical point TC$_{1}$ and TC$_{2}$ is rather strong, we have no understanding on why they should appear. From our construction of the Potts-3 and Potts-4 order parameter, it seems that the dimer degree of freedom is playing a dominating role in the low energy physics. It is thus attempting to write down an effective theory in terms of the dimer degree of freedom by integrating out other rapidly fluctuating degree of freedom. The origin of these two tricritical points may be more clearly seen from the perspective of such low energy effective theory. Interestingly, we find that the value of $J_{3}$ at TC$_{1}$($J_{3}\approx 0.42$) is very close to a transition point in the Schwinger Boson mean field phase diagram of the quantum $J_{1}-J_{3}$ KAFHM, where two nematic spin states with exactly the same symmetry but different PSG characters meet. 

The existence of ring structure in the low energy spin excitation spectrum will greatly enhance the fluctuation effect in two dimensional magnet\cite{Kane}. We thus expect the "grey region" to have good opportunity to realize some kind of quantum spin liquid phase when we consider the quantum version of the $J_{1}-J_{3}$ KAFHM. In the next section, we will present a Schwinger Boson mean field theory for the ground state phase diagram of the quantum $J_{1}-J_{3}$ KAFHM in the "grey region".

\section{V. A Bosonic RVB theory of the quantum $J_{1}-J_{3}$ KAFHM in the "grey region"}
\subsection{A. The Schwinger Boson mean field theory of a general quantum magnet}
The Schwinger Boson mean field theory(SBMFT) is widely used in the study of quantum magnet systems\cite{Arovas,Auerbach}. The SBMFT can describe the paramagnetic and the magnetic ordered phase of a quantum magnet on an equal footing and is thus especially convenient when we are concerned with quantum spin liquid state in close proximity to a magnetic ordered phase. Previous studies find that the SBMFT performs rather well in describing ground state property of spin-$\frac{1}{2}$ antiferromagnetic Heisenberg model on both the square and the triangular lattice\cite{Liang,Chen,Qiu}.

In the SBMFT, the spin operator is rewritten in terms of Bosonic spinon operator as 
\begin{equation}
\mathbf{S}_{i}=\frac{1}{2}\sum_{\alpha,\beta}b^{\dagger}_{i,\alpha}\bm{\sigma}_{\alpha,\beta}b_{i,\beta}.
\end{equation}
Here $\alpha,\beta=\uparrow,\downarrow$ is the spin index of the Bosonic spinon operator, $\bm{\sigma}$ is the usual Pauli matrix. The spinon operator should satisfy the constraint of 
\begin{equation}
\sum_{\alpha}b^{\dagger}_{i,\alpha}b_{i,\alpha}=1
\end{equation} 
to be a faithful representation of the spin algebra. In terms of the Schwinger Boson representation, the Heisenberg exchange coupling term can be written as
\begin{equation}
\mathbf{S}_{i}\cdot\mathbf{S}_{j}=\hat{B}^{\dagger}_{i,j}\hat{B}_{i,j}-\hat{A}^{\dagger}_{i,j}\hat{A}_{i,j},
\end{equation}
in which
\begin{eqnarray}
\hat{A}_{i,j}&=&\frac{1}{2}(b_{i,\uparrow}b_{j,\downarrow}-b_{i,\downarrow}b_{j,\uparrow})\nonumber\\
\hat{B}_{i,j}&=&\frac{1}{2}(b^{\dagger}_{i,\uparrow}b_{j,\uparrow}+b^{\dagger}_{i,\downarrow}b_{j,\downarrow}).
\end{eqnarray}
This form suggests the mean field decoupling of the Heisenberg exchange interaction in terms of the RVB order parameter $B_{i,j}=\langle\hat{B}_{i,j}\rangle$ and $A_{i,j}=\langle\hat{A}_{i,j}\rangle$. These set of RVB parameters are called an RVB mean field ansatz. To enforce the single occupancy constraint of the Bosonic spinon on average, we should also introduce the Boson chemical potential $\lambda$. This results in the Schwinger Boson mean field Hamiltonian of the following form
\begin{eqnarray}
H_{MF}&=&\sum_{i,j}J_{i,j}(B^{*}_{i,j}\hat{B}_{i,j}-A^{*}_{i,j}\hat{A}_{i,j})+\mathrm{H.C.}\nonumber\\
&+&\lambda\sum_{i,\alpha}b^{\dagger}_{i,\alpha}b_{i,\alpha}.
\end{eqnarray} 

\subsection{B. Determination of the RVB mean field ansatz for the $J_{1}-J_{3}$ KAFHM}
The key step in the SBMFT description of the ground state of a quantum magnet is the determination of the optimized value of the RVB parameters. 
For a quantum magnet with well-defined semiclassical ground state, an ansatz for the RVB parameter $A_{i,j}$ and $B_{i,j}$ can be easily obtained by looking at the semiclassical limit of the quantum model, in which we can take the spin operator $\mathbf{S}_{i}$ as a classical vector and the Boson operator $b_{i,\alpha}$ as a conventional complex number\cite{Qiu,Qiu1}. However, this approach fails badly in the "grey region" for the $J_{1}-J_{3}$ KAFHM. 

We are left only with the possibility of numerical optimization. In this approach, we write down a variational mean field Hamiltonian of the form of Eq.21 with the RVB parameter $A_{i,j}$, $B_{i,j}$ and the Boson chemical potential $\lambda$ treated as the variational parameters. We then calculate the variational ground state energy of the system as a function a these variational parameters and optimize it subjected to the average Boson occupation constraint. 

Such an optimization problem is in general very hard to solve. However, it can be significantly simplified when we enforce some symmetry requirement on the spin state we want to describe. Here we note that the Schwinger Boson representation of the spin operator Eq.17 has an intrinsic $U(1)$ gauge redundancy as a result of the no double occupancy constraint on the spinon operator(i.e., $\mathbf{S}_i$ is unchanged under the gauge transformation $b_{i,\alpha}\rightarrow b_{i,\alpha}e^{i\phi_{i}} $). Thus to describe a spin state with certain symmetry, the gauge non-invariant mean field RVB parameter $A_{i,j}$ and $B_{i,j}$ should be invariant under the symmetry operation only up to a $U(1)$ gauge transformation. It turns out that such symmetry conditions on the RVB mean field ansatz can be classified by the so called projective symmetry group(PSG) technique first developed by Wen in the Fermionic RVB theory\cite{psg}. It is then generalized to Bosonic RVB theory by some other authors\cite{psgb,psgc}. 

In our study, we will enforce translational symmetry on the spin state we want to describe. In the Schwinger Boson theory, the translational symmetry can be realized either by a translational invariant RVB mean field ansatz, or an RVB mean field ansatz with a doubled unit cell. In the latter situation, the translated ansatz differs from the original ansatz by a $U(1)$ gauge transformation. In the literature these two classes of RVB ansatz are called the uniform-flux and $\pi$-flux state. We find that throughout the phase diagram of the $J_{1}-J_{3}$ KAFHM the uniform-flux state always has a lower variational energy. We will focus on such a state in the following.

In the Schwinger Boson formalism, the mean field parameter $A_{i,j}$ and $B_{i,j}$ describe respectively antiferromagnetic and ferromagnetic local correlation. It is thus attempting to assume a non-zero $B_{i,j}$ between the first-neighboring sites and a nonzero $A_{i,j}$ between the third-neighboring sites for the $J_{1}-J_{3}$ KAFHM. However, in general we can also have first-neighboring pairing term $A_{i,j}$ and third-neighboring hopping term $B_{i,j}$.  Our numerical optimization indeed found such terms, although the first-neighboring hopping term and the third-neighboring pairing term are always the dominating variational parameters. By the way, our numerical optimization shows that the RVB parameters can always be chosen real. This is consistent with the absence of time reversal symmetry breaking in this model.

\subsection{C. SBMFT phase diagram of the $J_{1}-J_{3}$ KAFHM in the "grey region"}
If we only enforce the translational symmetry, then there will be 6 inequivalent first-neighboring bonds and 6 inequivalent third-neighboring bonds within each Kagome unit cell. We thus have 24 independent RVB parameters on these bonds. As we mentioned above, we find all these RVB parameters can be chosen real as a result of the time reversal symmetry. We are then left with 24 real variational parameters. Together with the Boson chemical potential we will have 25 real parameters to be optimized. The numerical optimization is done using the simulated annealing algorithm on a Kagome cluster with $8 \times 8$ unit cells. We first diagonalize the Schwinger Boson mean field Hamiltonian by Bogolyubov transformation at each of the $64$ momentum and then calculate the mean field ground state energy by the Wick decomposition of the model Hamiltonian, with the result\cite{Qiu}
\begin{equation}
E[A_{i,j},B_{i,j},\lambda]=-\frac{3}{8N_{s}}\sum_{i,j} J_{i,j} (\ |\langle \ \hat{A}_{i,j}\ \rangle |^{2}-| \langle \ \hat{B}_{i,j}\ \rangle |^{2}),
\end{equation}
in which $\langle \hat{A}_{i,j}\rangle$ and $\langle \hat{B}_{i,j}\rangle$ are the expectation value of the operator $\hat{A}_{i,j}$ and $\hat{B}_{i,j}$ defined in Eq.20 in the mean field ground state, $N_{s}$ denotes the number of lattice site in the cluster, which is $8\times8\times 3=192$ in our case. To enforce the average Boson number constraint, we supplement the variational ground state energy with a penalty function and optimize the following function
\begin{equation}
f=E[A_{i,j},B_{i,j},\lambda]+C\left( \frac{1}{N_{s}}\sum_{i,\alpha}\langle b^{\dagger}_{i,\alpha}b_{i,\alpha} \rangle-1 \right)^{2},
\end{equation}
in which $C$ is a large positive number. In our optimization we have set $C=100$.

\begin{figure}[h!]
\includegraphics[width=7cm,angle=0]{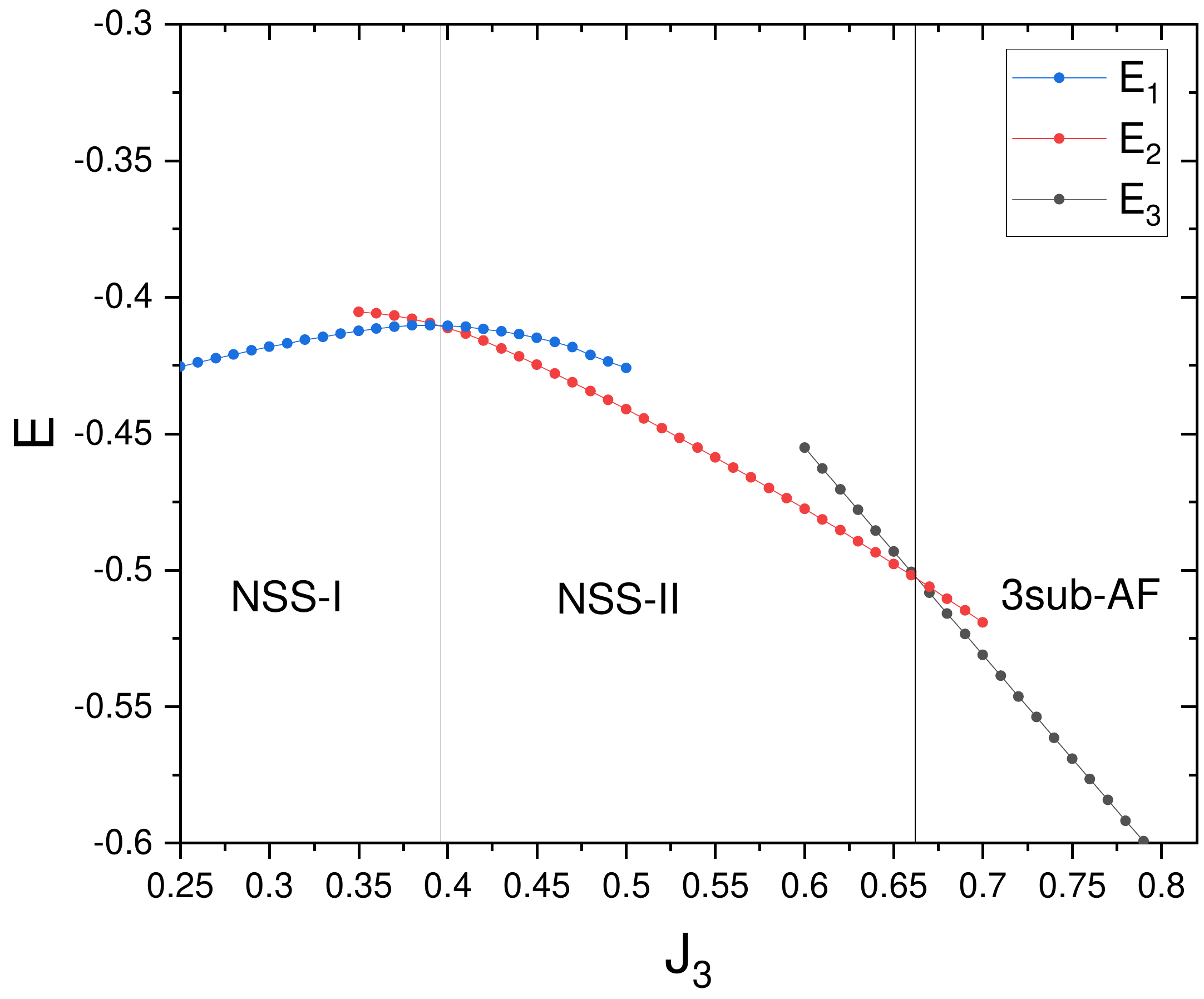}
\caption{The optimized Schwinger Boson mean field ground state energy of the $J_{1}-J_{3}$ KAFHM as a function of $J_{3}$. We find three solutions in the "grey region" from our simulated annealing optimization. These solutions are denoted as NSS-I, NSS-II and 3sub-AF respectively. } 
\label{fig11}
\end{figure}

The optimized Schwinger Boson mean field ground state energy of the $J_{1}-J_{3}$ KAFHM is plotted in Fig.11. We find three solutions in the "grey region"  from our simulated annealing optimization. These solutions are denoted as NSS-I, NSS-II and 3sub-AF respectively. Here NSS is short for nematic spin state, which breaks the three-fold rotation symmetry of the system. In fact, both NSS-I and NSS-II phase have the same $2mm$ symmetry(see Fig.2 for an illustration). The difference between them lies in the fact that their RVB mean field ansatz belong to different PSG classes. More specifically, the RVB mean field ansatz of the NSS-I phase is manifestly invariant under the action of $\sigma_{y}$ but requires an additional gauge transformation of the form 
\begin{equation}
\phi_{\sigma_{x}}(i)=\frac{\pi}{2}
\end{equation}
to recover its original form when acted with $\sigma_{x}$. On the other hand, the RVB mean field ansatz of the NSS-II phase is not invariant under the action of either $\sigma_{x}$ or $\sigma_{y}$. The gauge transformation required to recover the original form of the RVB mean field ansatz in the NSS-II phase is given by
\begin{eqnarray}
\phi_{\sigma_{y}}(i)=\left\{\begin{aligned}
                 \frac{\pi}{2} &\ \ \  &  for\  i\in  A \\  
                 \frac{\pi}{2} &\ \ \  & for \ i\in B \\  
                -\frac{\pi}{2} &\ \ \  & for \ i\in C
\end{aligned}  
\right.
\end{eqnarray}
for $\sigma_{y}$ and by
\begin{eqnarray}
\phi_{\sigma_{x}}(i)=\left\{\begin{aligned}
                 0 & \ \ \ &for \  i\in  A \\  
                0 & \ \ \ & for \ i\in B \\  
                \pi & \ \ \ &for \  i\in C
\end{aligned}  
\right.
\end{eqnarray}
for $\sigma_{x}$. Such a difference in the symmetry property of the mean field ansatz between the NSS-I and the NSS-II phase is best illustrated by the RVB parameters on the first-neighboring bonds(see Fig.12 and Fig.13). In particular, the first-neighboring pairing parameter in the NSS-I phase along the horizontal direction is identically zero as a result of the reflection symmetry of the RVB mean field ansatz under $\sigma_{y}$. On the other hand, the first-neighboring pairing parameter in the NSS-II phase along the horizontal direction is nonzero as a result of the nontrivial PSG of the RVB mean field ansatz under $\sigma_{y}$. The NSS-I phase thus has one less variational parameter than NSS-II phase(10 versus 11).

\begin{figure}[h!]
\includegraphics[width=7cm,angle=0]{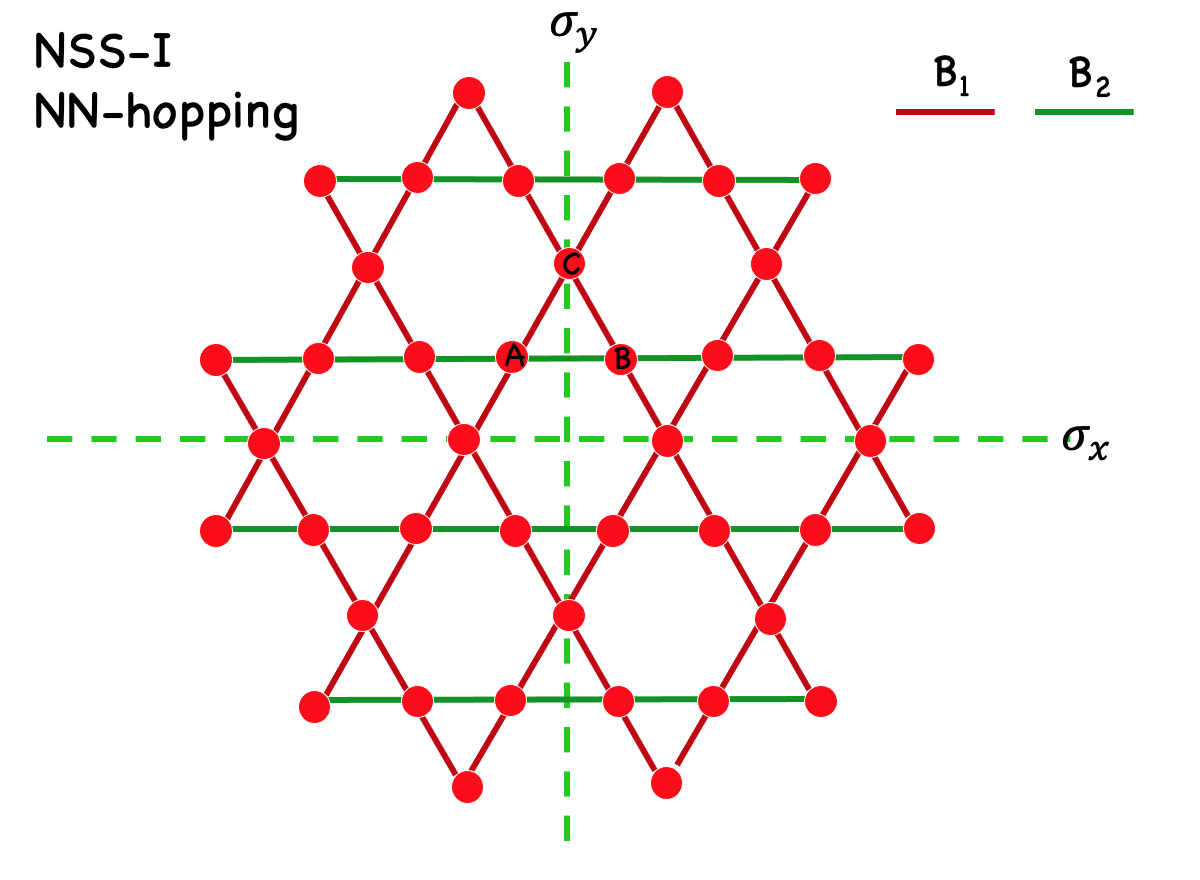}
\includegraphics[width=7cm,angle=0]{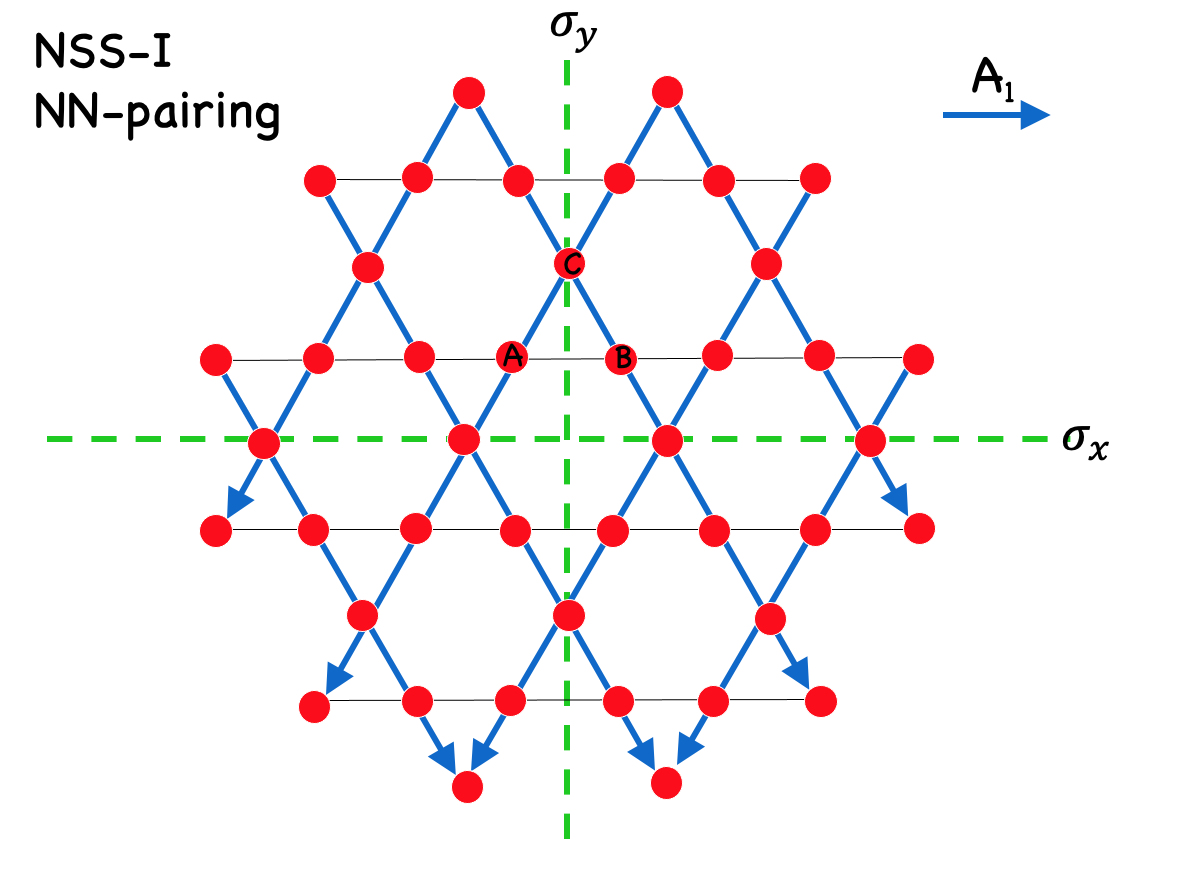}
\caption{Illustration of the RVB parameters on the first-neighboring bonds in the NSS-I phase. Note that the first-neighboring pairing parameter along the horizontal direction is zero as a result of the reflection symmetry of the RVB mean field ansatz under $\sigma_{y}$.} 
\label{fig12}
\end{figure}

\begin{figure}[h!]
\includegraphics[width=7cm,angle=0]{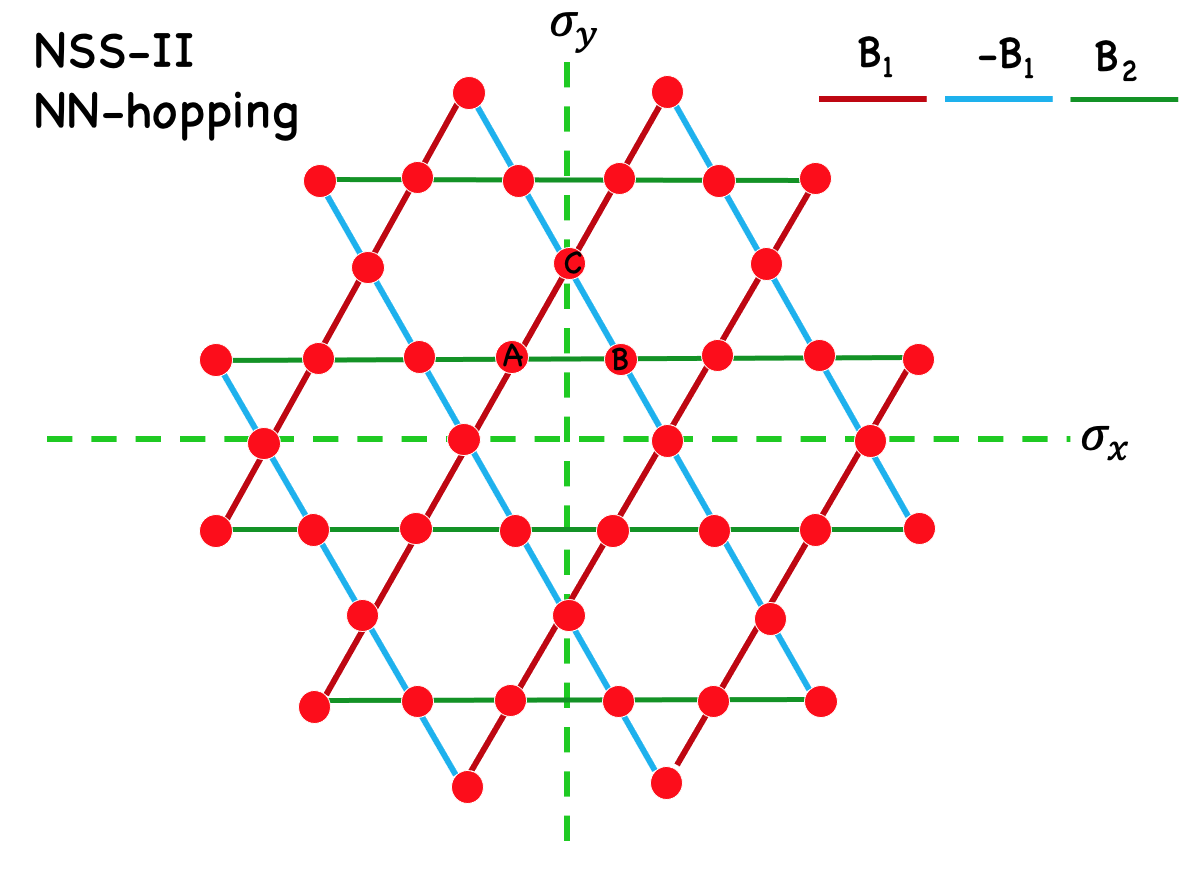}
\includegraphics[width=7cm,angle=0]{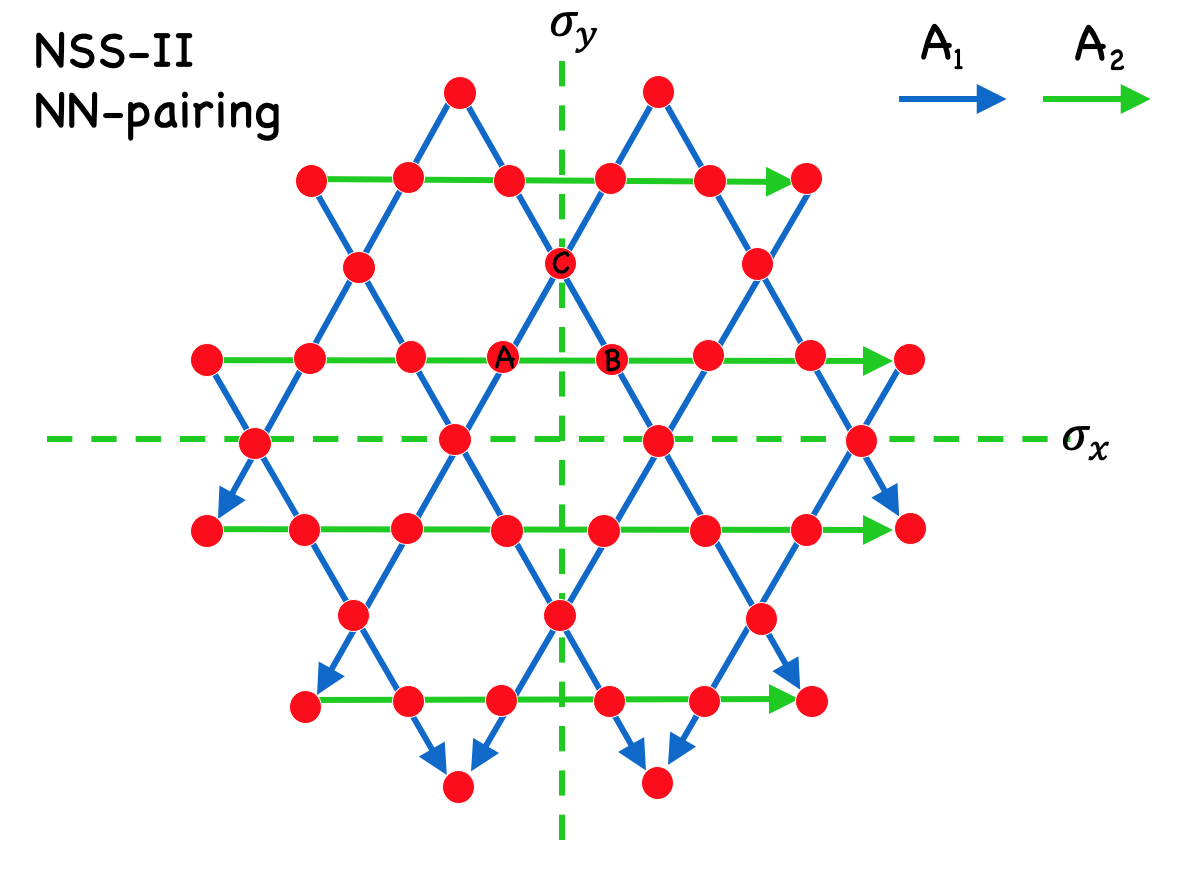}
\caption{Illustration of the RVB parameters on the first-neighboring bonds in the NSS-II phase. Note that the first-neighboring pairing parameter along the horizontal direction is now nonzero as a result of the nontrivial PSG of the RVB mean field ansatz under $\sigma_{y}$. } 
\label{fig13}
\end{figure}

Different from the NSS-I and the NSS-II phase, the 3sub-AF phase is fully symmetric. It is described by an RVB mean field ansatz with only third-neighboring pairing term, namely 
\begin{eqnarray}
               A_{i,j}=0,&  B_{i,j}=0 &  \mathrm{first \ neighbor}\nonumber \\  
               A_{i,j}=\pm A,&  B_{i,j}=0  &  \mathrm{third \ neighbor}.
\end{eqnarray}
Here the $\pm$ sign should be chosen so that no gauge flux is enclosed in any third-neighboring plaquette. The three sublattices are thus totally decoupled at the mean field level in this ansatz. Since the mean field ground state is independent of $J_{3}$ in the 3sub-AF phase, the optimized mean field ground state energy is found to be a strictly linear function of $J_{3}$.

\begin{figure}[h!]
\includegraphics[width=7cm,angle=0]{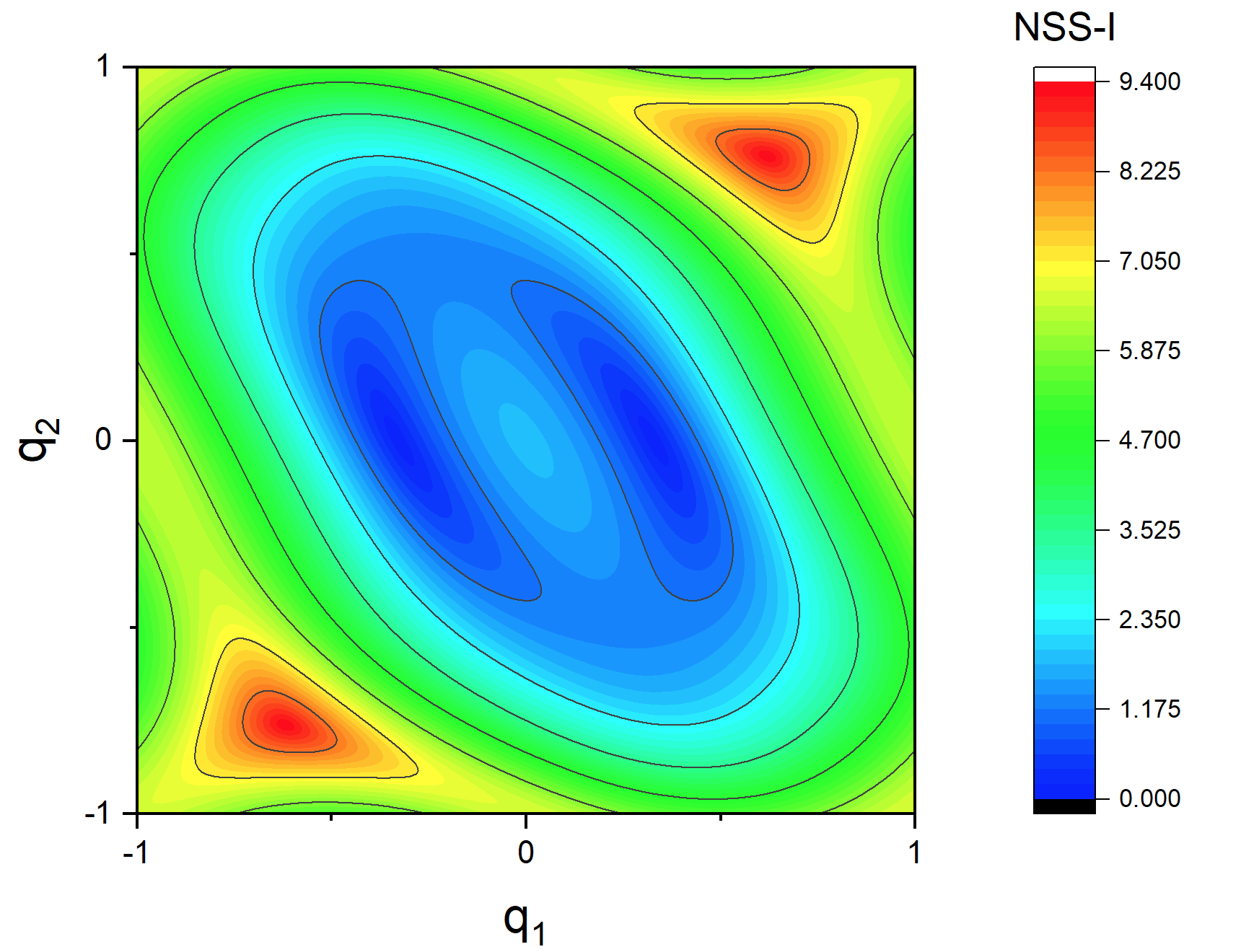}
\includegraphics[width=7cm,angle=0]{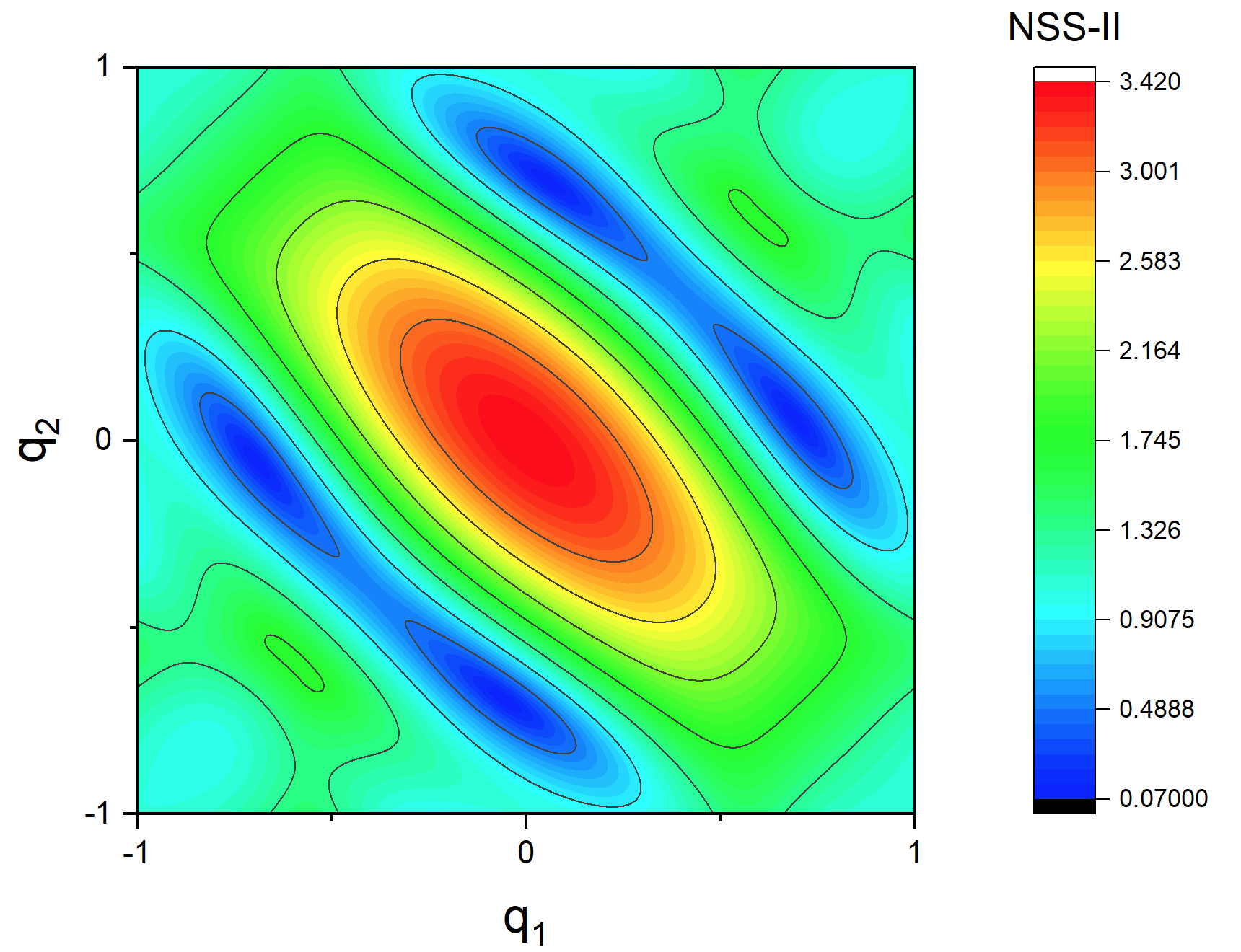}
\includegraphics[width=7cm,angle=0]{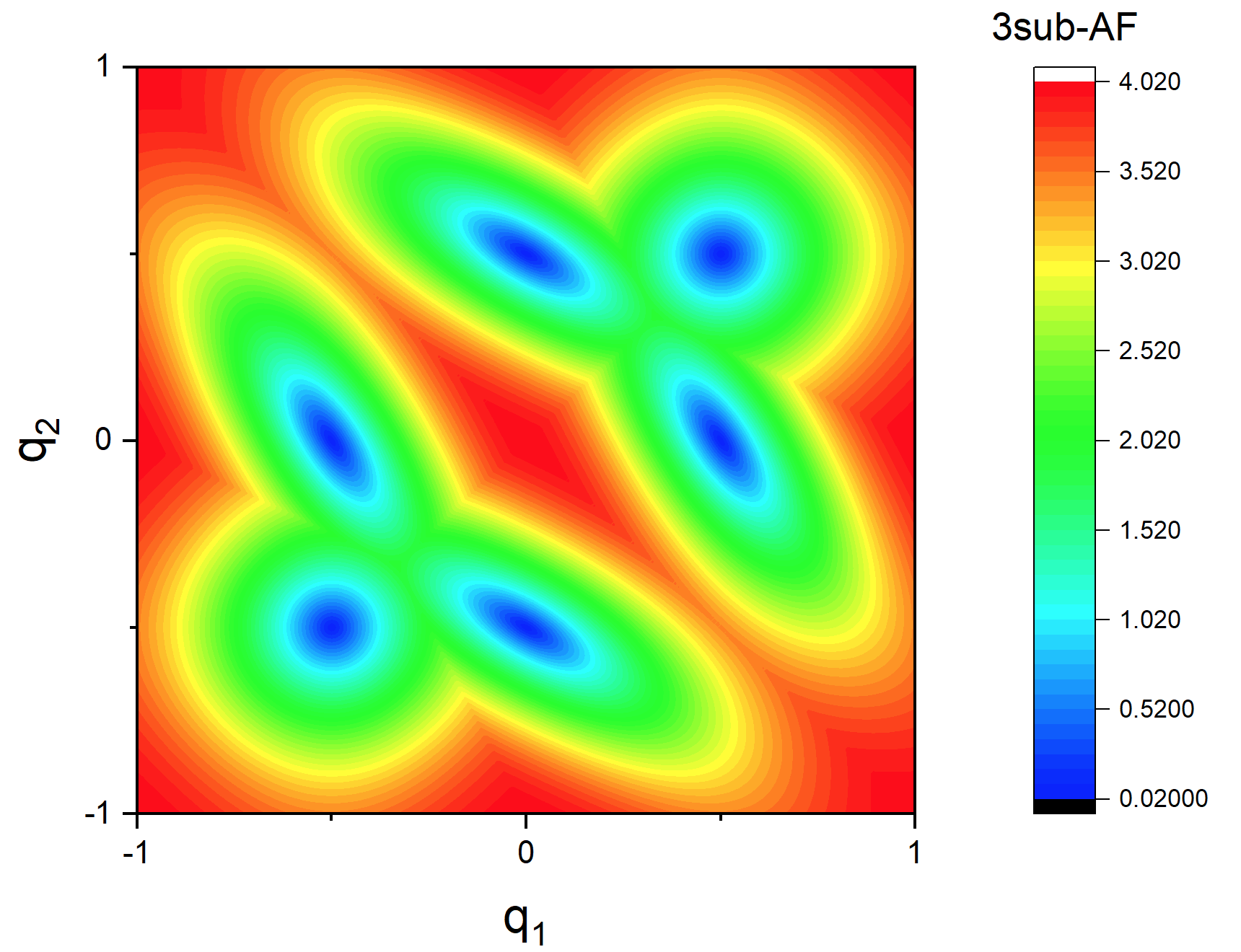}
\caption{Illustration of dispersion of the lowest spinon branch in the NSS-I, NSS-II and the 3sub-AF phase. The dispersion is calculated with the optimized RVB parameter at $J_{3}=0.35$(NSS-I), $J_{3}=0.5$(NSS-II) and $J_{3}=0.8$(3sub-AF). $q_{1}$ and $q_{2}$ are measured in unit of $\pi$.} 
\label{fig14}
\end{figure}

To have a better understanding on the physical difference between these phases, in particular, the difference between the NSS-I and the NSS-II phase, we have calculated the spinon dispersion within each phase. Shown in Fig.14 is the the dispersion of the lowest spinon branch in the NSS-I, NSS-II and the 3sub-AF phase. While the spinon dispersion in the NSS-II phase shows some similarity with that in the 3sub-AF phase, the spinon dispersion in the NSS-I phase, which features an anisotropic ring around the $\mathbf{q}=0$ momentum, is more close to that of the ferromagnetic phase. This explains why the NSS-I phase is favored over the NSS-II phase in the lower half of the "grey region".

We note that the transition point between the NSS-I phase and the NSS-II phase($J_{3}\approx 0.396$) is surprisingly close to the tricritical point TC$_{1}$ in the classical phase diagram of the model, which is located at $J_{3}\approx 0.42$. This is indeed intriguing if it is not a simple coincidence. It seems to imply an uncovered relation between the tricritical behavior in the classical model at finite temperature and the quantum phase transition behavior related to the change of quantum order(or PSG) in the ground state of the quantum model. It is interesting to see if similar behavior exist in more general frustrated magnetic models. 

At the same time, the phase boundary between the NSS-II phase and the 3sub-AF phase is very close to the phase boundary between the Potts-3 and the Potts-4 phase in the classical phase diagram of the model. This is not totally unexpected from the semiclassical limit of the Schwinger Boson representation. However, we note that the optimized mean field ansatz of the 3sub-AF phase describes three totally decoupled antiferromagnetic sublattices and does not possess the Potts-4 order. This may either imply the inadequacy of the Schwinger Boson mean field theory in describing such a phase, or the necessity to relax the translational symmetry on the mean field ansatz. This problem is left for future study. 

\subsection{D. Comparison with the exact diagonalization result on a 36-site cluster}
Motivated by the hope to find a quantum spin liquid state in the strongly frustrated "grey region", together with the intriguing closeness between the tricritical point TC$_{1}$ in the classical phase diagram and the transition point betwween the NSS-I and the NSS-II phase in the SBMFT phase diagram of the quantum model, we have carried out an exact diagonalization study of the ground state of the quantum $J_{1}-J_{3}$ KAFHM on a 36-site cluster with the geometry illustrated in Fig.15. We have used the Lanczos algorithm to calculate the lowest eigenvalue of the spin-$\frac{1}{2}$ $J_{1}-J_{3}$ KAFHM within the fully symmetric subspace, which contains 31,527,894 basis vectors.

\begin{figure}[h!]
\includegraphics[width=7cm,angle=0]{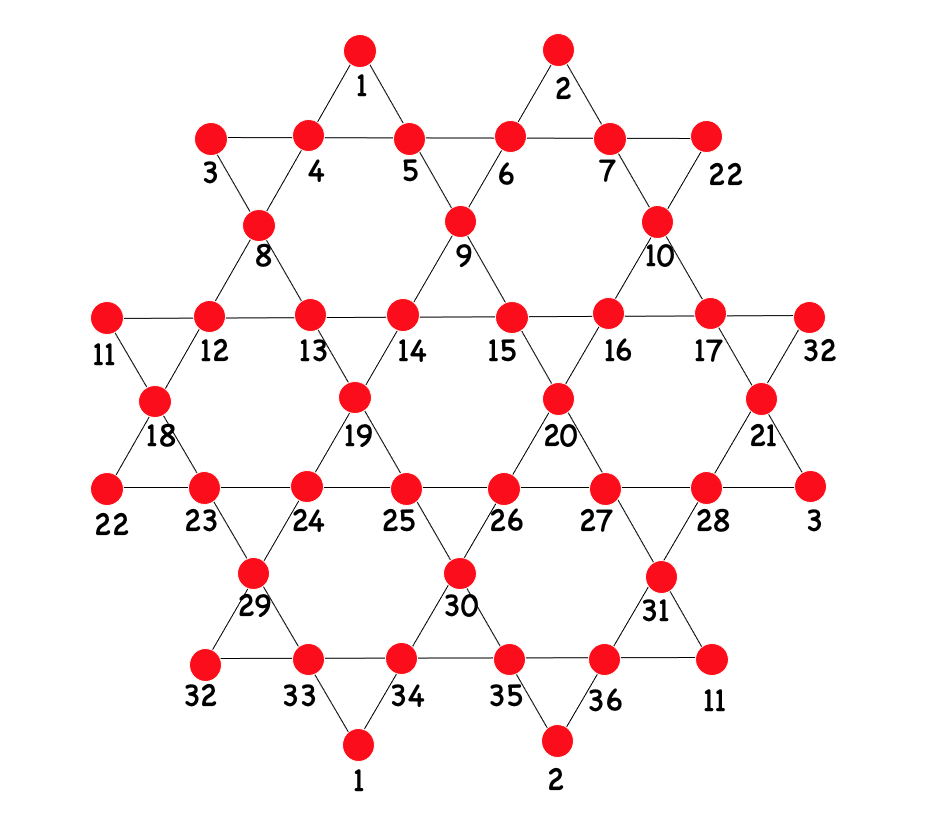}
\caption{The 36-site cluster on which our exact diagonalization calculation is performed.} 
\label{fig15}
\end{figure}

The result of the ground state energy as a function of $J_{3}$ is shown in Fig.16. The ground state phase diagram of the model on the 36-site cluster is characterized by two transitions at $J_{3}\approx 0.28$ and $J_{3}\approx 0.42$. For $J_{3}< 0.28$, the ground state of the system is found to be the fully polarized ferromagnetic state. The ground state energy as a function of $J_{3}$ is thus strictly linear in this region. For $J_{3}> 0.42$, the ground state of the system evolves continuously into the ground state of the $J_{3}$-only model, which consists three idecoupled sublattices of antiferromagnetic correlated spins. We find that the energy gain from inter-sublattice coupling is quite substantial even for $J_{3}=0.8$(as high as $4.5\%$). The quantum fluctuation correction can thus substantially enhance the stability of the 3sub-AF phase. This may explain why the region of the intermediate phase is substantially suppressed as compared to the Potts-3 phase in the classical phase diagram and the nematic spin state in the SBMFT phase diagram. In particular, the NSS-II phase in the SBMFT phase diagram may be totally swallowed by the fluctuation-reinforced 3sub-AF phase.

\begin{figure}[h!]
\includegraphics[width=7cm,angle=0]{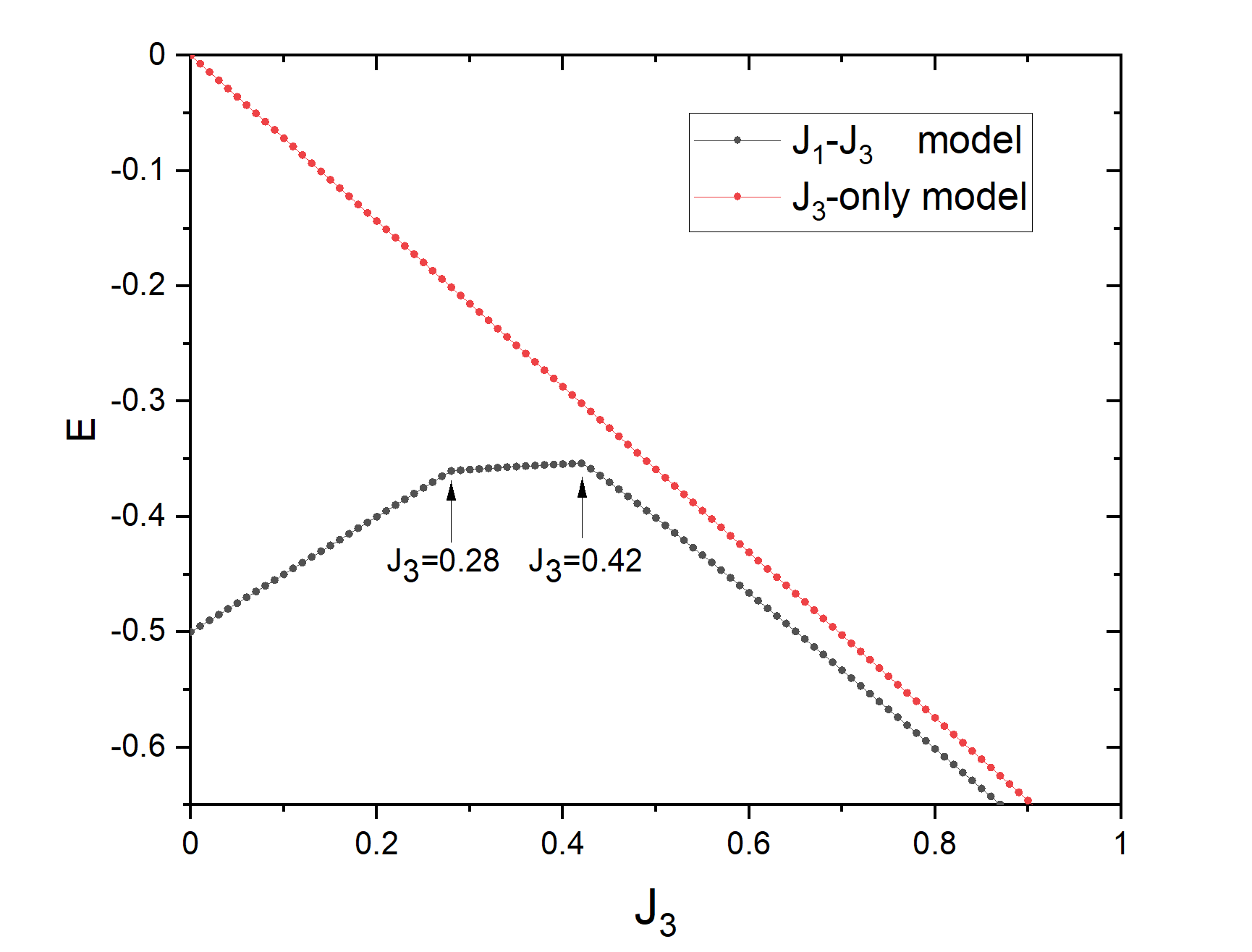}
\caption{The ground state energy of the spin-$\frac{1}{2}$ $J_{1}-J_{3}$ KAFHM as a function of $J_{3}$ calculated on the 36-site cluster shown in Fig.15. The red line is the result when we discard the inter-sublattice coupling $J_{1}$. The energy gain from the quantum fluctuation correction is quite substantial even for rather large $J_{3}$.} 
\label{fig16}
\end{figure}

We note that the transition point between the 3sub-AF phase and the intermediate phase is surprisingly close to the vanishing point of the NSS-I phase in the SBMFT phase diagram, and is even closer to the tricritical point TC$_{1}$ in the classical phase diagram of the $J_{1}-J_{3}$ KAFHM. We think such closeness in the phase boundaries may not be simple coincidence. Of course, calculation on cluster of larger size or with different geometry is needed to check if Fig.16 indeed represent the intrinsic behavior of the quantum $J_{1}-J_{3}$ KAFHM. A tensor network simulation of the model can also be very helpful to resolve the situation\cite{TRG2}. 

A comparison of Fig.16 with the classical phase diagram shown in Fig.3 and the SBMFT phase diagram shown in Fig.11 strongly suggests the identification of the intermediate phase in Fig.16 as the nematic spin state NSS-I. Such a spin state is characterized by a Potts-3 order and a spinon dispersion with an anisotropic ring structure around the $\mathbf{q}=0$ momentum(see Fig.14). Such an anisotropic ring structure will also be inherited by the spin structure factor of the system. This phase can actually be understood as an anisotropic generalization of the double spiral phase proposed in Ref.[\onlinecite{Kane}], in which the low energy fluctuation along the ring in the spinon dispersion can easily disorder the long range correlation in the spin channel. With these considerations in mind, we think it is very possible that the $J_{1}-J_{3}$ KAFHM can realize a nematic spin liquid state in the intermediate region of $J_{3}$. It is interesting to see if a nematic spin liquid with the same quantum order as the NSS-I phase can be verified in more accurate study of the model in the future.

\section{VI. Discussions}
In this work, we have mapped out the complete classical phase diagram of the $J_{1}-J_{3}$ KAFHM through extensive Monte Carlo simulation. We find that in the most strongly frustrated region of the model, namely the "grey region" ($\frac{1}{4} \leq J_{3}\leq \frac{1+\sqrt{5}}{4}$) in which the Luttinger-Tizsa criteria fails to predict the classical ordering pattern of the system, competing $q=3$ Potts order and $q=4$ Potts order emerge. We find that while the $q=4$ Potts phase evolves continuously into the 3sub-AF phase in the large $J_{3}$ limit, the Potts-4 order in general can not be understood as the consequence of the conventional order-by-disorder effect with preexisting antiferromagnetic ordered sublattices. This becomes especially clear if one note how deeply the Potts-4 phase has penetrated into the "grey region". In fact, we find no evidence for any phase transition at the upper boundary of the "grey region". The Potts-4 phase meets the Potts-3 phase within the "grey region" with a first order transition between the two. We thus feel it is better to take the Potts-4 order, rather than the antiferromagnetic correlation within each of the three sublattices of the Kagome lattice as the most  fundamental structural signature of this phase. The emergence of the Potts-3 order, which occurs entirely within the "grey region", is certainly beyond the description of the order-by-disorder mechanism. We think it is the strong preference of the system to resolve the frustration in the "grey region" that drive the emergence of both the Potts-3 and the Potts-4 order in this region. 

Our simulation indicates that while the Luttinger-Tisza criteria fails to predict the phase boundary of the system in the strongly frustrated "grey region", it is very useful in predicting the behavior of the spin structure factor of the system. On general ground, this is just what we should expect from the Luttinger-Tisza criteria since the spontaneous breaking of the spin rotational symmetry at finite temperature is prohibited by the Mermin-Wagner theorem in two dimension systems. We find that the spin structure factor of the system in the "grey region" is characterized by some anisotropic ring structure centered either around $\mathbf{q}=0$, or around the three momentum $\mathbf{q}_{1}=(\pi,0)$,  $\mathbf{q}_{2}=(0,\pi)$ and $\mathbf{q}_{3}=(\pi,\pi)$. The existence of such ring structure will greatly enhance the fluctuation effect and encourage the emergence of exotic phases at both classical and quantum level.    
  
Tricritical points are found along the phase boundary of both the Potts-3 and the Potts-4 phase in the classical phase diagram of the $J_{1}-J_{3}$ KAFHM. It is almost impossible to see why these tricritical points should emerge from the behavior of the spin degree of freedom directly. To understand the mechanism by which these tricritical point emerge, a low energy effective description of the system in terms of the dimer degree of freedom seems to be necessary. Such an effective description may be achieved by a careful integration of the fast fluctuating single spin degree of freedom.

Our Schwinger Boson mean field theory calculation confirms the existence of the Potts-3 order in the ground state of the quantum $J_{1}-J_{3}$ KAFHM in the "grey region". However, such a nematic spin state is found to take two different forms distinguished by their projective symmetry group character(or quantum order), although they have exactly the same symmetry. The nematic spin state in the lower half of the "grey region", namely NSS-I, is found to support a spinon dispersion featuring an anisotropic ring structure around the $\mathbf{q}=0$ and can be understood as the anisotropic generalization of the double spiral phase proposed for doped t-J model three decades ago. We find that the transition point between the NSS-I state and the NSS-II state is intriguingly close to the tricritical point TC$_{1}$ in the classical phase diagram of the model. Further work to check if this is just a simple coincidence is obviously of interest. 

Our Schwinger Boson mean field theory fails to produce the Potts-4 order, even though the predicted phase boundary between the NSS-II phase and the 3sub-AF phase is very close to the phase boundary between the Potts-3 and the Potts-4 phase in the classical phase diagram. This may either imply the inadequacy of the Schwinger Boson mean field theory in describing such a phase, or the necessity to relax the translational symmetry on the mean field ansatz. A further study on this issue is obviously necessary.

Finally, we have compared the classical phase diagram and the Schwinger Boson mean field phase diagram with the result obtained from exact diagonalization on a 36-site cluster. It turns out that the quantum fluctuation correction from first-neighboring exchange coupling can substantially reinforce the 3sub-AF phase. However, an intermediate phase does exist in between the 3sub-AF phase at higher $J_{3}$ and the ferromagnetic phase at lower $J_{3}$. We propose this intermediate phase to be a nematic spin liquid phase with the same quantum order as the NSS-I phase. Intriguingly again, the intermediate phase is found to terminate at a point very close to the tricritical point TC$_{1}$ in the classical phase diagram of the model. Further numerical study is obviously needed to decide if such a nematic spin liquid state indeed exist and if there is indeed any deep implications in all these intriguing closeness in phase boundaries.

In all, we find that rich competition of emergent orders of classical or quantum nature can occur in the "grey region" of the $J_{1}-J_{3}$ KAFHM, in which the system lost its preferred semiclassical ordering pattern. We think this a general trend in frustrated magnets. The "grey region" generated by strong frustration effect should be better taken as an opportunity to discover more exotic phases, rather than a burden on theoretical analysis on such frustrated models.

We acknowledge the support from the National Natural Science Foundation of China(Grant No. 11674391), the Research Funds of Renmin University of China(Grant No.15XNLQ03), and the National Program on Key Research Project(Grant No.2016YFA0300504) and the open project of key laboratory of artificial structures and quantum control(Ministry of Education), Shanghai Jiaotong University. We thank Tsuyoshi OKUBO for the help on the over relaxation technique in Monte Carlo simulation.


\begin{thebibliography}{99}
\bibitem{dimer}In principle, higher order quantities or dimers of longer range can also be used to describe the finite temperature transition of the $J_{1}-J_{3}$ KAFHM. However, from symmetry point of view, except for some accidental situation the first-neighboring dimer suffices for such a purpose. Our numerical simulation shows that the first-neighboring dimer may even acts as the primary order in both the 3sub-AF phase and the mystery "grey region".
\end{thebibliography}
\end{document}